\documentclass[authoryear,12pt]{elsarticle}

\usepackage{amsmath}
\usepackage{amssymb}
\usepackage{graphicx}
\usepackage{lineno}

\journal{}

\begin{document}
	
	\begin{frontmatter}
		
		
		\title{Time-dependent numerical model for simulating internal oscillations in a sea organ}
		
		\author[gradri]{Nino Krvavica\corref{mycorrespondingauthor}}
		\cortext[mycorrespondingauthor]{Corresponding author}
		\ead{nino.krvavica@uniri.hr}
		\author[indeko,gradri]{Gabrijel Peroli}
		\author[gradri]{Igor Ru\v{z}i\'{c}}		
		\author[gradri]{Nevenka O\v{z}ani\'{c}}		
		
		\address[gradri]{University of Rijeka, Faculty of Civil Engineering, Radmile Matejcic 3, 51000 Rijeka, Croatia}
		\address[indeko]{IND-EKO d.o.o., Korzo 40, 51000 Rijeka, Croatia}

		\begin{abstract}
			This paper presents a one-dimensional time-dependent numerical model of a sea organ, which generates music driven by the motion of the sea. The governing equations are derived by coupling hydrodynamic and thermodynamic equations for water level and air pressure oscillations in a sea organ pipe system forced by irregular waves. The model was validated by comparing numerical results to experimental data obtained from a scaled physical model. Furthermore, the models' capabilities are presented by simulating internal oscillations in the Sea Organ in Zadar, Croatia. The response of the Sea Organ varies between segments and for different wave conditions. The strongest air pressure and water level response is found near resonance frequencies. 
		\end{abstract}
		
		
	\end{frontmatter}

\section{Introduction}

A sea organ is an acoustical, architectural and hydraulic structure, which uses the motion of the sea to generate music. The original idea dates back to 3rd century BC when a so-called \textit{hydraulis} was invented by Ctesibius of Alexandria \citep{britannica}. This mechanical pipe organ consisted of several acoustical pipes placed on top of a wind chest that was connected to a wind chamber. The sound was produced by a compressed air flowing through the pipes. The wind chamber was half filled with water so that when the air pressure decreased, pumps were manually activated to increase the water level, which compressed the air and restored the required pressure in the wind chest. 

\begin{figure}[th]
	\center
	\includegraphics[width=7.9cm]{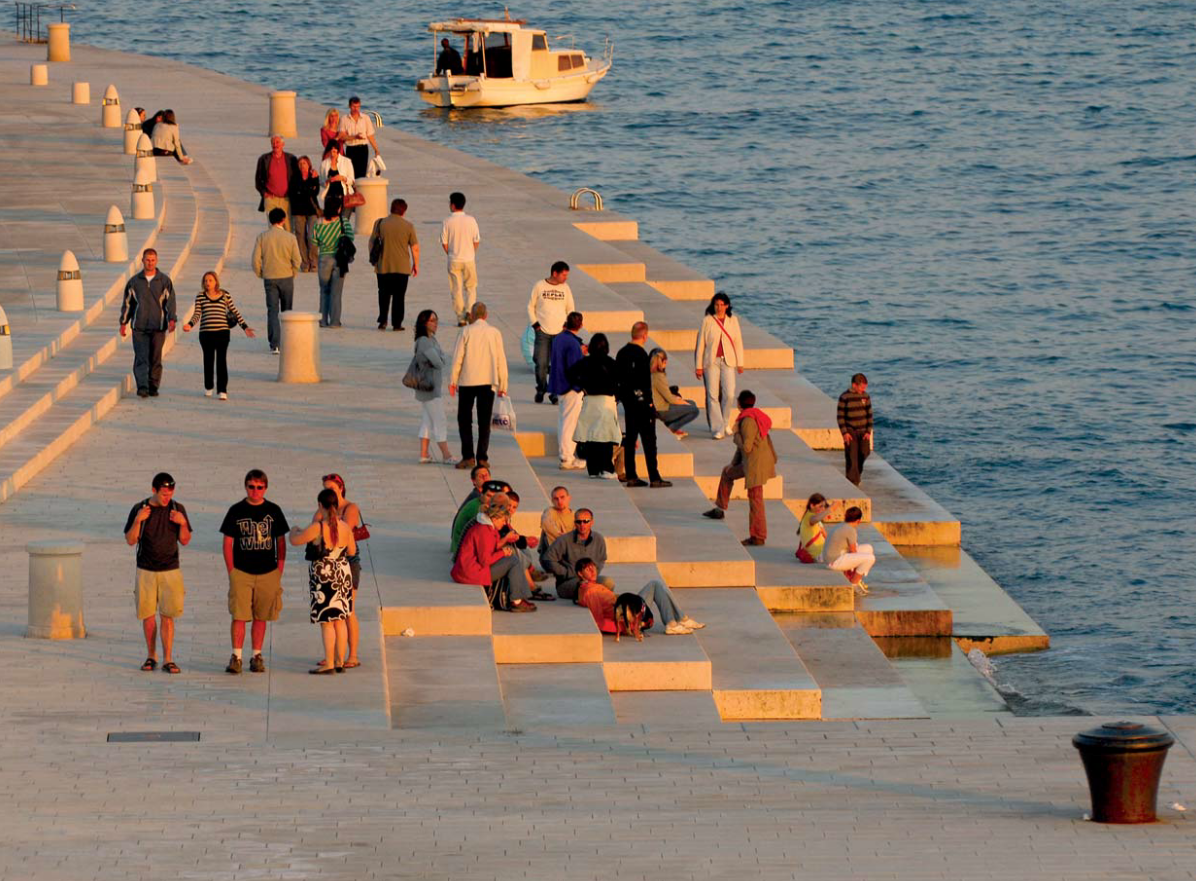}
	\hfill
	\includegraphics[width=5.63cm]{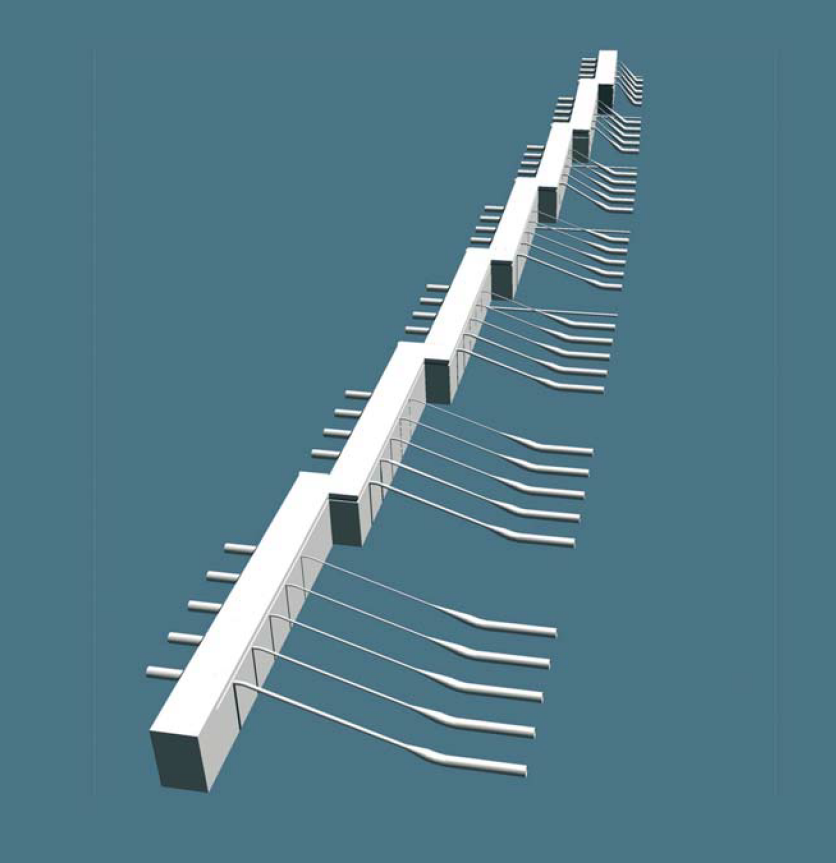}
	\vfill
	\caption{Photograph of the Sea Organ in Zadar and a 3D model of the organ pipe system}
	\label{fig:sea_organ}
\end{figure}

This idea was reinvented in the 1980's by constructing the Wave Organ in the San Francisco Bay \citep{waveorgan}. The Wave Organ uses the stochastic motion of waves and tides to compress the air in the pipes and generate random sounds. The Sea Organ, on the other hand, is as much a musical instrument as it is a complex coastal and hydraulic structure. It was designed by Nikola Ba\v{s}i\'{c} and opened to the public in 2005 in Zadar, Croatia \citep{stelluti2011}. This 75-m long instrument was built by reconstructing a deteriorated sea-wall at the Zadar promenade. On the outside, the structure is defined by seven segments of stone steps descending into the sea (Figure \ref{fig:sea_organ}). But underneath those steps, each segment contains five organ pipes of various lengths and diameters specifically constructed to produce notes of a certain frequency (Figure \ref{fig:sea_organ}). 

In recent years, the Sea Organ has become one of the most popular tourist attraction in Croatia, and it has received numerous international awards \citep{stelluti2011,rossetti2011}. Its acoustical and musical characteristics have been thoroughly analysed and presented to a scientific community \citep{stamac2005,stamac2007,kapusta2007}. However, its hydraulic aspects are equally intriguing but have not yet been properly examined. It should also be noted that the design and construction of the Sea Organ were largely experimental due to lack of available numerical models at the time that could accurately simulate the complex multiphase hydrodynamic processes. 

A first attempt at simulating the hydraulic and musical aspects of the Sea Organ was a simplified computational model presented recently by \cite{krvavica2017sea}. This integrated approach consists of a computational algorithm for generating random waves, a one-dimensional (1D) numerical model for simulating the water level oscillations inside the pipes and a conceptual model for generating the sound. The numerical model was based on the assumption of negligible air compressibility and a linear relationship between the internal water level oscillations and air velocity in the acoustical pipe. This simplification is reasonable for relatively large openings, such as air ducts or some turbines \citep{koo2010}. However, preliminary experiments on a scaled sea organ model \citep{peroli2017} indicated that the air compressibility is significant enough to affect the internal water mass oscillations in the Sea Organ.

This paper presents a modified and extended numerical model that can simulate non-linear and time-dependent oscillations of both water level and air pressure in any sea organ. The proposed model is derived by coupling 1D hydrodynamic and thermodynamic equations, which describe the internal oscillations driven by the motion of the sea surface. This approach is based on similar studies for simulating wave energy converters, namely oscillating water columns (OWC) \citep{gervelas2011,iino2016}, but with differently defined hydrodynamic equations due to a more complex geometry. The proposed model is validated by comparing the computed and experimental results obtained from a scaled physical model. 

The paper is organized as follows; first, the hydraulic characteristics of the Sea Organ are examined and described; next, the time-dependent numerical model is derived and presented; also, the experimental set-up is shown; and finally, the results of model validation and analysis of the Sea Organ in Zadar are presented and discussed, followed by the conclusion.

\section{Hydraulic characteristics of the Sea Organ}

\begin{figure}[thbp]
	\center
	\includegraphics[width=14cm]{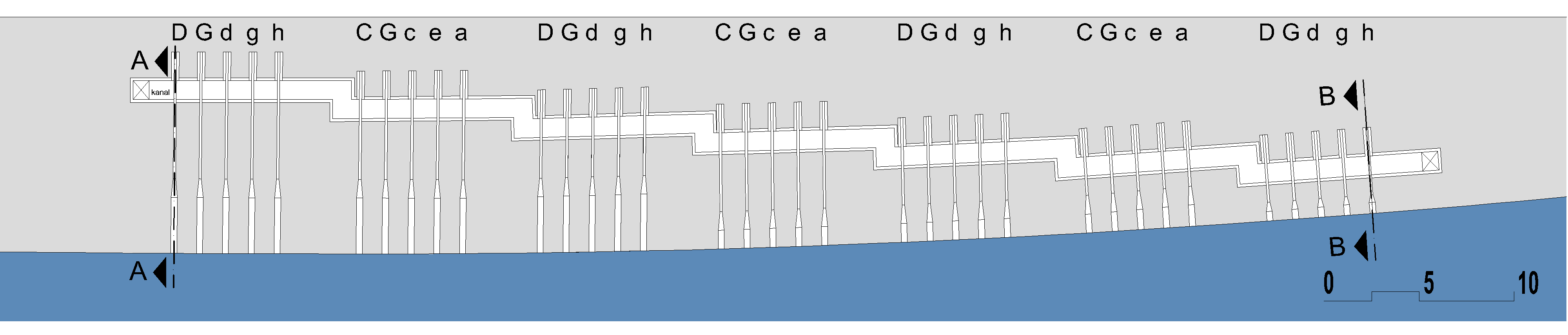}
	\vfill
	\includegraphics[width=14cm]{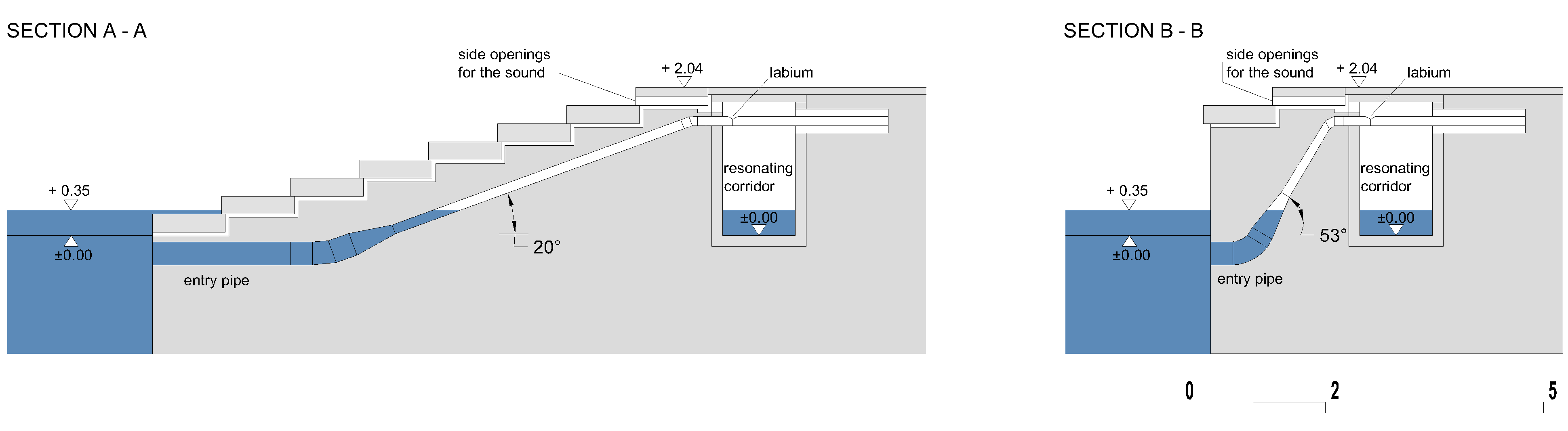}
	\caption{Characteristic plan and longitudinal sections of the Sea Organ in Zadar}
	\label{fig:seaorgan_design}
\end{figure}

The Sea Organ in Zadar is a 75-m long structure divided into seven segments. Each segment contains five organ pipes of various lengths and diameters, and each pipe consists of three distinct parts  (Fig.~\ref{fig:seaorgan_design}):
(\textit{i}) the first (entry) pipe of a larger diameter is submerged below the sea surface and positioned horizontally, (\textit{ii}) the second (sloped) pipe of a smaller diameter is positioned on an inclined surface facing upwards, and (\textit{iii}) the third (acoustical) pipe is positioned horizontally under the walking surface. The first two pipes are made from polyethylene (PE), whereas the third pipe is made of stainless steel, it is closed at the end but has a small orifice at the beginning \citep{stelluti2011}.

The processes of generating the sound is quite simple; the waves, tides, and passing boats initiate the movement of the sea surface; the vertical movement of the sea surface in front of the sea organ forces the water level oscillations inside the pipes; the internal water mass then compresses the air pushing it through the acoustical pipe, where a sound of a predefined frequency is finally produced. The sound emanates from the top and side openings in the steps. In this way, nature itself determines the duration and intensity of each note, but the arrangement of the pipes, each tuned to a different frequency, governs the resulting melody.

 From a musical point of view, every odd segment in specifically tuned to produce five tones from a G-major chord (D-G-d-g-h), whereas every even segment produces five tones from C-major chord with additional sixth (C-D-c-e-a), as illustrated in Figure \ref{fig:seaorgan_design} \citep{stamac2005,stamac2007}. All tones correspond to frequencies in the range 60-250 Hz. However, to achieve the required sound wave frequency, the dimensions of the labium and the resonant pipe must be designed accordingly. Since the dimension of the labium orifice governs the air velocity, it may also affect the water level and air pressure oscillations in the pipes.

\section{Time-dependent numerical model}

The numerical sea organ model is developed by combining a hydrodynamic model for water mass oscillations and a thermodynamic model for the air pressure variations. First, the governing equations for each model are derived. Next, the coupling between these equations is presented. And finally, the numerical scheme for solving the governing system of equations is presented. The algorithm code has been implemented in Python 3.6.

\subsection{Hydrodynamic governing equations}

Governing equations for water mass oscillations in organ pipes are derived from the law of conservation of mass and energy for incompressible and irrotational fluid.
The integral form of the mass conservation law for a control volume (CV) bounded by a control surface (CS) is written as \citep{white1999}:
\begin{equation}
\frac{\textrm{d}}{\textrm{d} \textmd{t}} \int_{CV} \rho \textrm{d}V
+ \int_{CS} \rho \textbf{u} \cdot \textbf{n} \textrm{d}A
= 0,
\label{eq:con_mass}
\end{equation}
where the first term denotes the mass rate of change inside the CV, and the second term denotes the mass flux across CS, also d$V$ is an element volume, d$A$ is an element area of the control surface, $t$ is time, $\rho$ is the fluid density, \textbf{u} is the fluid velocity vector (with components $u,v,w,$), and \textbf{n} is a unit vector normal and directed outwards from the control surface at any point.

Similarly, the energy conservation law for CV may be written as \citep{white1999}:
\begin{equation}
\frac{\textrm{d}}{\textrm{d} \textmd{t}} \int_{CV} e \rho \textrm{d}V
+ \int_{CS} e \rho (\textbf{u} \cdot \textbf{n}) \textrm{d}A = - \dot{W} =
- \int_{CS} p (\textbf{u} \cdot \textbf{n}) \textrm{d}A
-\dot{W_f},
\label{eq:con_energy}
\end{equation}
where the first term denotes the energy rate of change inside the CV, the second term denotes the energy flux across CS, and the right-hand side denotes the work done by the system. In the present study, the work done by the pressure and the shear work due to viscous stresses (friction) was considered. Also, $e = gz + (u^2 + v^2 + w^2)/2$ is the system energy per unit mass, $g$ is the acceleration of gravity, $z$ is the elevation, and $p$ is the pressure.

For a pipe element with no other inflow or outflow other than its entry point and under the assumption of constant density, both mass and energy conservation equations can be reduced to one dimension. With velocity and energy per unit mass averaged over the pipe cross-section area, Eq.~(\ref{eq:con_mass}) is rewritten as follows:
\begin{equation}
\frac{\textrm{d}V}{\textrm{d}t} = Q, 
\label{eq:continuity}
\end{equation}
where $Q$ is the volumetric flow rate in the pipe, respectively. Furthermore, Eq.~(\ref{eq:con_energy}) is divided by the mass flow rate $\rho Q$ and acceleration of gravity $g$, and rewritten in dimensions of length:
\begin{equation}
\frac{1}{g} \int_{1}^{2} \frac{\textrm{d} \textbf{u}}{\textrm{d} t} \textrm{d}l = 
H_{1} - H_{2} - \Delta H,
\label{eq:dynamic}
\end{equation}
where $l$ is the length of the pipe along its axis between entry-point 1 and endpoint 2, $\Delta H$ is the energy dissipation represented in terms of a head loss, and $H_{i}$ is the total head that accounts for the potential and kinetic energy, as well as the pressure at some point $i$ along the pipe axis:
\begin{equation}
H_{i} = z_{i} + \frac{p_{i}}{\rho g} + \frac{\alpha Q^2}{2 g A_{i}^2},
\label{eq:total_head}
\end{equation}
where $z_{i}$ is elevation, $\alpha$ is the Coriolis coefficient (kinetic energy correction factor) and $A_{i}$ is the cross-section area of the pipe at any point $i$. Note, that the frictionless form of Eq.~(\ref{eq:dynamic}) is identical to the unsteady Bernoulli's equation. However, viscosity and friction are an important aspect of internal processes and they should not be omitted from the governing equations. 

\begin{figure}[thbp]
	\center
	\includegraphics[width=12cm]{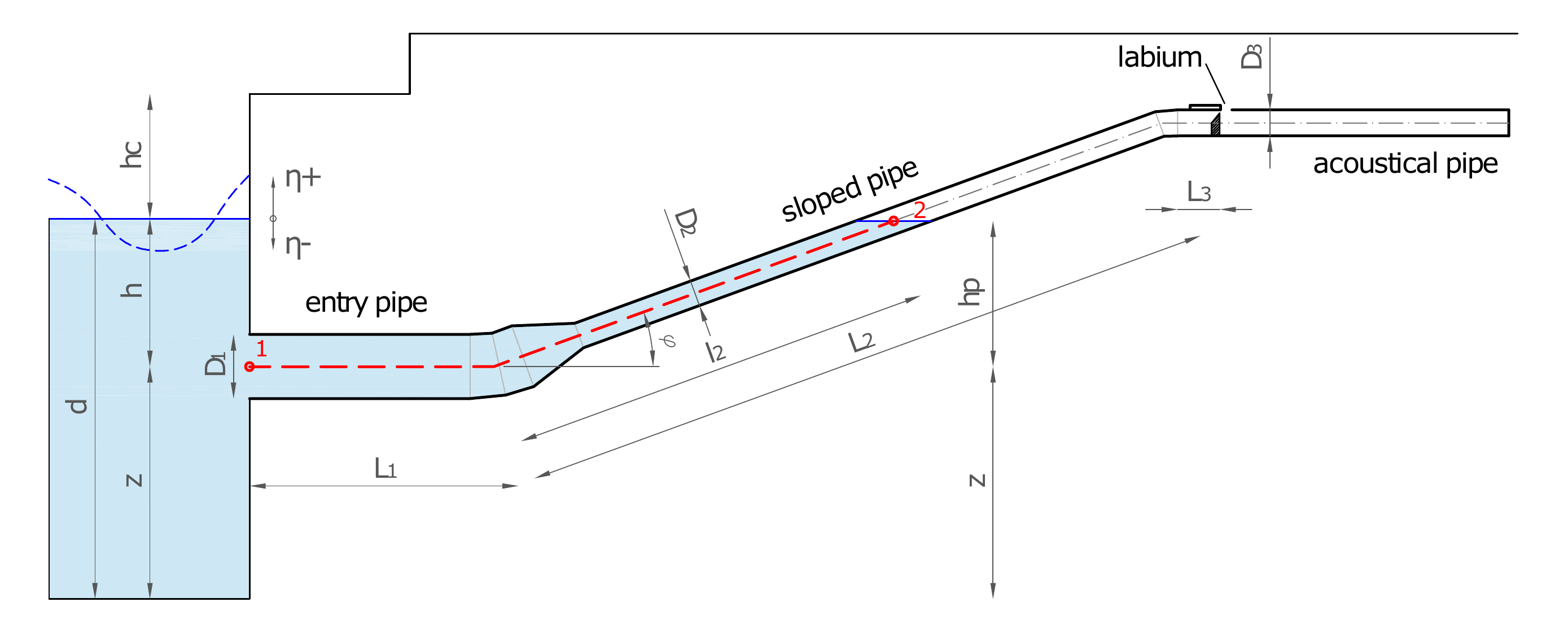}
	\caption{One-dimensional numerical model scheme of the organ pipe system}
	\label{fig:scheme}
\end{figure}

Let us now consider a special case of a sea organ pipe system that consists of three connected pipes of variable sizes, as described in the previous section and illustrated in Fig.~\ref{fig:scheme}. Under the assumption that the water level is always position somewhere along the second pipe, Eq.~(\ref{eq:continuity}) may be rewritten as follows:
\begin{equation}
\frac{\textrm{d}l_2}{\textrm{d}t} = \frac{Q}{A_2},
\label{eq:continuity2}
\end{equation}
where $l_2$ is the length of the water column along the second pipe axis (see Fig.~\ref{fig:scheme}), $Q$ is the volumetric flow rate of the water in the pipe system, and $A_2$ is the cross-section area of the sloped pipe.

The energy equation (\ref{eq:dynamic}) is modified as follows: (\textit{i}) the term on the left hand side is integrated along the first two pipes (from the pipe entry (1) to the water level (2)), (\textit{ii}) term $H_{1}$ is replaced by the total wave-induced pressure head $p_{wave}/(\rho g)$ at the depth $h$ corresponding to the centre of the pipe entry, (\textit{iii}) energy dissipation is accounted for by minor and major head losses, which are then collected and defined in terms of the coefficient $\beta$ and kinetic energy. To ensure the correct sign of energy dissipation terms, the kinetic energy is written as a product of the flow rate $Q$ and its absolute value $\vert Q\vert$. Finally, the energy equation for the water oscillations in a sea organ pipe system is given by:
\begin{equation}
\left( \frac{L_1}{gA_1} + \frac{l_2}{gA_2}\right) \frac{\textrm{d} Q}{\textrm{d} t} = 
\frac{p_{wave}}{\rho g} - \frac{\Delta p}{\rho g} - l_2 \sin \varphi  - \beta \frac{Q\vert Q\vert}{2g},
\label{eq:dynamic2}
\end{equation}
where $L_1$ is the length of the first pipe, $A_1$ is the cross-section of the first pipe, $\Delta p = p - p_{atm}$ is the air pressure drop inside the organ pipe (difference between the absolute air pressure $p$ and atmospheric pressure $p_{atm}$), $\varphi$ is the inclination angle between the axis of the sloped pipe and the horizontal plane, and $\beta$ is defined as:
\begin{equation}
\beta = \frac{\alpha}{A_2^2} + \frac{\xi_E}{A_1^2} + \frac{\xi_A}{A_1^2} + \frac{\xi_R}{A_2^2} + \lambda_1 \frac{L_1}{D_1 A_1^2} + \lambda_2 \frac{l_2}{D_2 A_2^2},
\label{eq:beta}
\end{equation}
where $\xi_E$ is the loss coefficient at the pipe inlet, $\xi_A$ is the loss coefficient at the pipe elbow, $\xi_R$ is the loss coefficient duo to profile reduction, the pipe friction losses are defined by the Darcy-Weisbach equation \citep{white1999}, where $D_{1,2}$ are the diameters for the respective first and second pipe, and $\lambda_{1,2}$ are the respective friction coefficients, usually computed by the implicit Colebrook-White equation or its explicit approximation \citep{haaland1983}. Simple expressions for all of these coefficients are well-known and readily available in most classical books on hydraulics or fluid mechanics, \textit{e.g.}, \citep{white1999}.

\subsection{Water wave pressure}

The wave pressure $p_{wave}$ at the pipe inlet is computed by the linear wave theory \citep{sorensen1993}. Let us first consider regular harmonic wave that propagates in the $x$ direction:
\begin{equation}
	\eta(x,t) = a \cos \left( \omega t - kx + \phi \right),
	\label{eq:harmonic_wave}
\end{equation}
where $\eta$ is the surface elevation, $a$ is the wave amplitude, $\omega$ is the wave angular frequency, $k$ is the wave number, and $\phi$ is the wave phase. Given wave height $H$ and length $L$, these parameters can be determined from simple relations: $a=H/2$, $k=2\pi/L$, and $\omega=\sqrt{gk \tanh(kd)}$, where $d$ is the total water depth. The wave pressure under a regular wave at some depth $h$ is defined by hydrostatic and hydrodynamic components \citep{sorensen1993}:
\begin{equation}
	p_{wave} = p_{stat} + p_{dyn} = \rho g h + \rho g \eta(x,t) \frac{\cosh\left[k(d-h)\right]}{\cosh(kd)}.
\end{equation}

However, to account for the randomness of real waves and describe their stochastic nature, the irregular surface elevation at a given distance $x$ are computed here by a random phase-amplitude model based on a spectral description of wind-generated waves \citep{holthuijsen2010,krvavica2017sea}. This is implemented in the proposed algorithm by computing the sum of a finite number of harmonic waves, defined by different wave amplitudes and phases, as follows:
\begin{equation}
	\eta(x,t) = \sum_{i=1}^{N} \eta_i(x,t) = \sum_{i=1}^{N} a_i \cos \left[ \omega_i t - k_i x + \phi_i \right],
\end{equation}
where $N$ is a finite number of spectral components (denoted by index $i$). Each harmonic wave has a unique amplitude $a_i(\omega_i) = \sqrt{2S_{\eta}(\omega_i) \Delta \omega}$, which is derived from a given wave density spectrum $S_{\eta}(\omega_i)$ discretized by a finite number of frequency increments $\Delta \omega = \omega_{max} / N$. Usually, the Pierson-Moskowitz \citep{pierson1964} or JONSWAP spectrum \citep{hasselmann1973} is used for this purpose; although, the T-spectrum \citep{tabain1997} has proven to be more realistic for the Adriatic Sea \citep{parunov2011}. Also, each wave has a unique phase $\phi_i$ which is randomly selected from a uniform distribution. 

The wave reflection from the sea-organ wall may also be accounted for by a local increase in the wave amplitude. Therefore, the wave amplitude near the organ sea-wall is locally redefined as:
\begin{equation}
	a = a_{in} + a_{ref} = (1 + K_r) a_{in},
\end{equation}
where $a_{in}$ is the incident wave amplitude, $a_{ref}$ is the reflected wave amplitude, and $K_r$ is the reflection coefficient, which is computed based on the crest height $h_c$ above the sea water level as follows \citep{goda2000}:
\begin{equation}
	K_r = \frac{6}{11} \frac{h_c}{H_s} + 0.7,
\end{equation}
where $H_s$ is the significant wave height.

Finally, the wave pressure under irregular waves at depth $h$ can be computed by the following expression:
\begin{equation}
	p_{wave} = \rho g h + \rho g (1+K_r) \sum_{i=1}^{N} a_i \frac{\cosh\left[k_i(d-h)\right]}{\cosh(k_i d)} \cos \left[ \omega_i t - k_i x + \phi_i \right].
	\label{eq:wave_pressure}
\end{equation}

\subsection{Thermodynamic governing equations}

An additional equation for the air pressure in the organ pipe is derived based on the thermodynamic principles. According to the ideal gas law, the air pressure $p$ is related to the gas density $\rho$ and temperature $T$ as follows \citep{white1999}:
\begin{equation}
	p = \rho RT,
	\label{eq:ideal_gas}
\end{equation}
where $R$ is the specific gas constant. 

Similarly to OWCs, the process of periodic compression and expansion of the air in the organ pipe can be considered as reversible and adiabatic, \textit{i.e.}, isentropic \citep{sarmento1985}. Similar assumption has proven to be justified in various OWCs \citep{josset2007,gervelas2011,iino2016}. Therefore, under the assumption of constant specific heat, the air pressure and temperature are related as follows \citep{white1999}:
\begin{equation}
Tp^{(\gamma-1)/\gamma} = \textmd{const.}
\label{eq:isentropic}
\end{equation}
where $\gamma$ is the heat capacity ratio ($\gamma = 1.4$ for air). 

The equation for the rate of change of pressure is obtained by differentiating Eq.~(\ref{eq:ideal_gas}) over time \citep{gervelas2011}, which gives:
\begin{equation}
\frac{\textrm{d} p}{\textrm{d} t} = 
\rho R \frac{\textrm{d} T}{\textrm{d} t} + RT \frac{\textrm{d} \rho}{\textrm{d} t}.
\label{eq:diff_ideal_gas}
\end{equation}
Next, Eq.~(\ref{eq:isentropic}) is also differentiated in respect to time, which gives:
%
\begin{equation}
\frac{\textrm{d} T}{\textrm{d} t} = \frac{\gamma - 1}{\gamma} \frac{T}{p} \frac{\textrm{d} p}{\textrm{d} t}.
\label{eq:dTdt}
\end{equation}

Inserting Eq.~(\ref{eq:dTdt}) into Eq.~(\ref{eq:diff_ideal_gas}) gives:
\begin{equation}
\frac{\textrm{d} p}{\textrm{d} t} = 
\frac{\gamma - 1}{\gamma} \frac{\rho R  T}{p} \frac{\textrm{d} p}{\textrm{d} t} + RT \frac{\textrm{d} \rho}{\textrm{d} t}.
\label{eq:dVdt_dpdt}
\end{equation}
Using Eq.~(\ref{eq:ideal_gas}) and after some algebraic manipulation, Eq.~(\ref{eq:dVdt_dpdt}) can be simplified to:
\begin{equation}
\frac{\textrm{d} p}{\textrm{d} t} = \frac{\gamma p}{\rho} \frac{\textrm{d} \rho}{\textrm{d} t}.
\label{eq:dpdt_0}
\end{equation}
Considering that the gas density changes in time due to temporal changes of mass and volume, Eq.~(\ref{eq:dpdt_0}) is finally written in the form:
\begin{equation}
	\frac{\textrm{d} p}{\textrm{d} t} = \frac{\gamma p}{m} \frac{\textrm{d} m}{\textrm{d} t}
	 - \frac{\gamma p}{V} \frac{\textrm{d} V}{\textrm{d} t} .
	\label{eq:dpdt}
\end{equation}

Time derivative of the air mass inside a pipe can be expressed as a negative mass flow rate $\dot{m}$ through the labium orifice \citep{wylie1993,gervelas2011}:
\begin{equation}
	\dot{m} = \textrm{sign}(\Delta p) C_d A_0 \sqrt{2 \vert \Delta p \vert \rho_{air} },  
	\label{eq:massflowrate}
\end{equation}
where $\Delta p$ is the air pressure drop, $C_d$ is the discharge coefficient and $A_0$ is the area of the labium orifice. Value for $C_d$ is usually determined experimentally, and it ranges from 0.4 to 0.7 \citep{lingireddy2004}. 
By inserting Eq.~(\ref{eq:massflowrate}) in Eq.~(\ref{eq:dpdt}), the following equation is obtained:
\begin{equation}
		\frac{\textrm{d} p}{\textrm{d} t} = 
		\frac{\textrm{d} \Delta p}{\textrm{d} t} =
		\textrm{sign}(\Delta p) \frac{\gamma p C_d A_0 }{m}  \sqrt{2 \vert \Delta p \vert \rho_{air} }
		- \frac{\gamma p}{V} \frac{\textrm{d} V}{\textrm{d} t} .
		\label{eq:ddpdt}
\end{equation}

For the organ pipe (Fig.~\ref{fig:scheme}), the time derivative of the volume of air can be expressed as the volumetric flow rate of water inside the pipe system. The volume of air inside the organ pipes may be computed as $V = A_3L_3 + A_2(L_2-l_2)$, where $A_3$ and $L_3$ are the respective cross-section area and length of the acoustical pipe. Furthermore, the speed of sound is introduced, which for the ideal gas may be defined as $c^2 = \gamma p / \rho$ \citep{wylie1993}. Finally, Eq.~(\ref{eq:ddpdt}) is simplified to:
\begin{equation}
	\frac{\textrm{d} \Delta p}{\textrm{d} t} = 
	\frac{\textrm{sign}(\Delta p) c^2 C_d A_0 \sqrt{2 \vert \Delta p \vert \rho_{air} }
	- \gamma p Q}{A_3L_3 + A_2(L_2-l_2)}.
	\label{eq:dpdt_organ}
\end{equation}

\subsection{The governing system of equations for a sea-organ pipe system}
 
Considering both hydrodynamic and thermodynamic processes presented in previous subsections, the problem of simulating water level and air pressure oscillations in a sea organ is defined by coupling three first-order ordinary differential equations. Equations (\ref{eq:continuity2}) and (\ref{eq:dynamic2}) define the oscillatory motion of the internal water level, whereas the third equation (\ref{eq:dpdt_organ}) defines the air pressure oscillations. The governing system of equations is defined as follows:
\begin{equation}
	\begin{cases}
		\dfrac{\textrm{d}l_2}{\textrm{d}t} = \dfrac{Q}{A_2} \\
		\dfrac{\textrm{d} Q}{\textrm{d} t} = 
		\dfrac{p_{wave}/\rho - \Delta p/\rho - g l_2 \sin \varphi  - \beta Q\vert Q\vert/2}{ L_1/A_1 + l_2/A_2 } \\
		\dfrac{\textrm{d} \Delta p}{\textrm{d} t} = 
		\dfrac{\textrm{sign}(\Delta p) c^2 C_d A_0 \sqrt{2 \vert \Delta p \vert \rho_{air} }
		- \gamma p Q}{A_3L_3 + A_2(L_2-l_2)}
		\label{eq:governing_eq}
	\end{cases}
\end{equation}
where three unknowns are the length of the water column in the second pipe $l_2$, volumetric flow rate of the water $Q$, and air pressure drop $\Delta p$. These processes are strongly coupled and codependent; therefore, the equations must be solved simultaneously.


\subsection{Numerical scheme}

A most common approach for solving any dynamical system is the direct numerical integration \citep{lambert1973}. This approach is based on satisfying a numerical approximation of the governing system of equations at discrete points in time, with a given initial solution. Many numerical methods, whether explicit or implicit, are available for this purpose. In this work, the implicit trapezoidal rule \citep{lambert1973} was applied to numerically evaluate the governing system of equations (\ref{eq:governing_eq}). 

The proposed trapezoidal rule for solving any ODE of the form
\begin{equation}
\frac{\textrm{d} y}{\textrm{d} t} = f(t,y)
\end{equation}
is defined as follows \citep{lambert1973}:
\begin{equation}
y^{n+1} = y^n + \dfrac{\Delta t}{2} \left[ f\left(t^n,y^n\right) + f\left(t^{n+1},y^{n+1}\right) \right]
\end{equation}
where superscript $n$ denotes known values at previous time step and $n+1$ denotes unknown values at time $t^{n+1}=t^n+\Delta t$, where $\Delta t$ is the time step. The trapezoidal rule is second-order accurate and A-stable numerical method \citep{lambert1973}. 

However, the method is implicit for non-linear equations and, therefore, some iterative method must be used. Since analytical formulation for the Jacobian matrix of the governing system is non-trivial (mainly due to derivatives of the empirical friction equation), a quasi-Newton method is preferred. The Broyden method \citep{broyden1965} was chosen here to solve the system of equations (\ref{eq:governing_eq}). This iterative method is based on replacing the Jacobian matrix by a discrete approximation, which is then easily updated at each iterative step (see \citep{broyden1965} for more details). Furthermore, the amount of computations at each step is reduced, and the convergence is superlinear \citep{broyden1965}.

\section{Laboratory experiments}

\begin{figure}[thbp]
	\center
	\includegraphics[width=12cm]{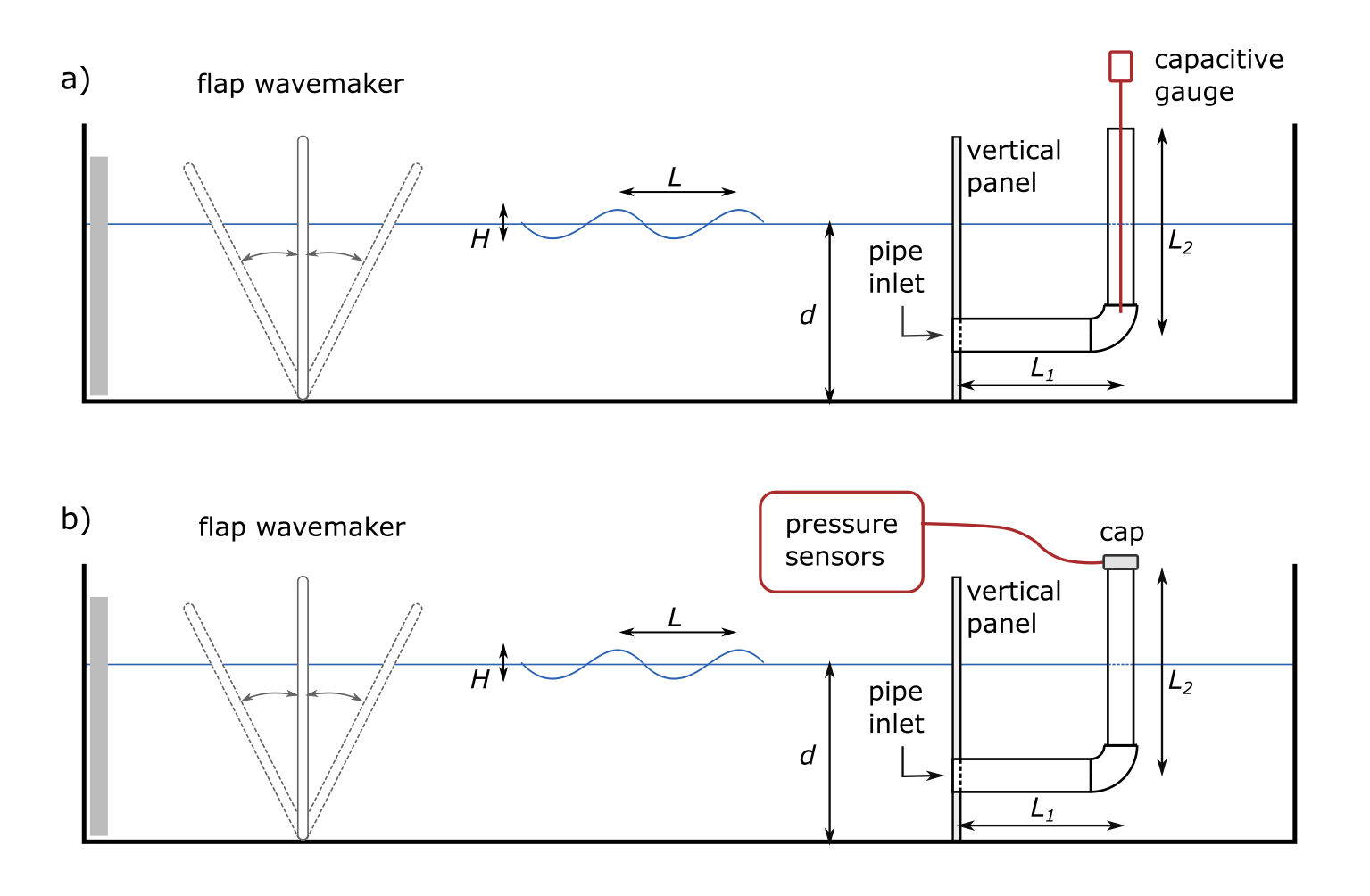}
	\caption{Scheme of the laboratory experiment for: A) open-end pipe, B) closed pipe with a perforated cap (not in scale)}
	\label{fig:exp_Scheme}
\end{figure}

To validate the proposed model, several experiments were conducted in the Hydraulic Laboratory at the University of Rijeka. An approximate 1:5 scale model of a sea organ pipe system was constructed in a 12.5 m long wave flume. The model set-up is illustrated in Fig.~\ref{fig:exp_Scheme}. 

The sea organ model consisted of a vertical panel (representing a sea wall), with a perforated round opening near the bottom, which was connected to an L-shaped organ-like pipe system. The first pipe, made out of PE with the inner diameter $D_1=32$ mm, was positioned horizontally and was connected by a $90^{\circ}$ elbow to the vertical pipe, made out of acrylic glass (PMMA) with $D_2 = 26$ mm. Three different lengths of the horizontal pipe were tested, $L_1= 20$, 40 and 60 cm. The length of the vertical pipe was $L_2 = 55$, 65 and 65 cm, respectively. 

Furthermore, to account for the influence of the labium orifice area on the air pressure drop, a plastic cap was installed at the top of the vertical pipe. Three caps with different orifice area were 3D printed, namely $A_0 = 1\times 6$, $1 \times 12$ and $1\times 18$ mm$^2$. The acoustical pipe was left out to simplify the construction of the physical model; however, Eq.~(\ref{eq:governing_eq}) still applies when $L_3=0$.

The water depth was set to $d=30$ cm, and the centre of the pipe inlet was positioned at 10 cm height from the bottom. A flap-type wavemaker was used to generate regular waves. Wavemaker paddle frequency was varied in the range from 0.3 to 0.7 Hz, which produced waves of different heights and lengths. Wavemaker equation may be used here to predict regular wave heights \citep{krvavica2017wave}. In general, wave generator set to a higher frequency produced higher waves at the same depth and paddle stroke. However, because of the reflection from the vertical panel and from the wavemaker paddle, spurious waves appeared, which were especially noticeable at lower frequencies. Therefore, after some time, the generated wave field, consisted of quasi-regular periodic waves. 


Two scenarios were considered: (\textit{i}) free-surface water mass oscillations (open-end pipes as illustrated in Fig.~\ref{fig:exp_Scheme}A) and (\textit{ii}) compressed-air water mass oscillations (partially closed pipes by a perforated cap as illustrated in Fig.~\ref{fig:exp_Scheme}B). In both cases, the total pressure under the wave was measured at the pipe inlet (at $h=20$ cm). In the first scenario, water elevations in the vertical pipe were measured by a capacitive gauge; however, this was not possible when the pipes were closed by a cap, therefore, only the air pressure drop under the cap was measured.

\section{Results}

The validation of the proposed model against experimental values is presented, as well as the numerical analysis of the Sea Organ in Zadar under different wave conditions.

\subsection{Model validation}

To validate the proposed model, numerical results are compared to measured values for systems with open-end pipes and for pipe systems closed by a perforated cap.

\subsubsection{Free surface water mass oscillations}

The experiment was set up as described in the previous section and illustrated in Fig.~\ref{fig:exp_Scheme}A. The parameters for the numerical model were defined based on the experiments' dimensions, and a constant air pressure, $\Delta p(t) = 0$. Therefore, only the first two expressions in Eq.~(\ref{eq:governing_eq}) were active. The wave pressure measured at the pipe inlet was imposed as the boundary condition for the numerical model. 

\begin{figure}[thbp]
\center
\includegraphics[width=6.7cm]{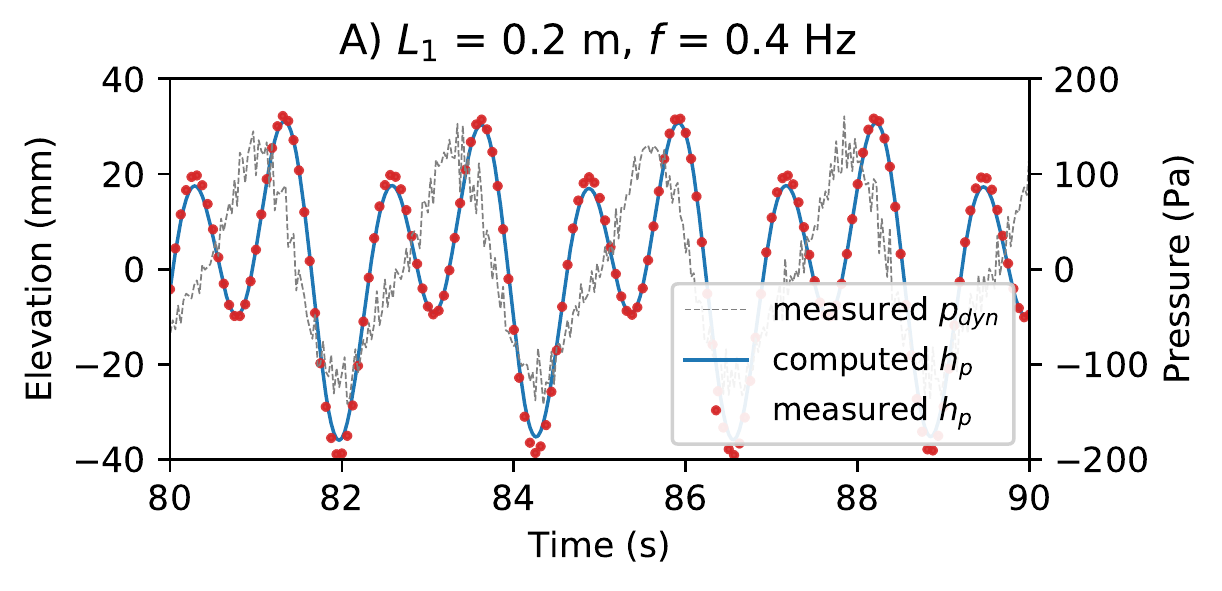}
\hfill
\includegraphics[width=6.7cm]{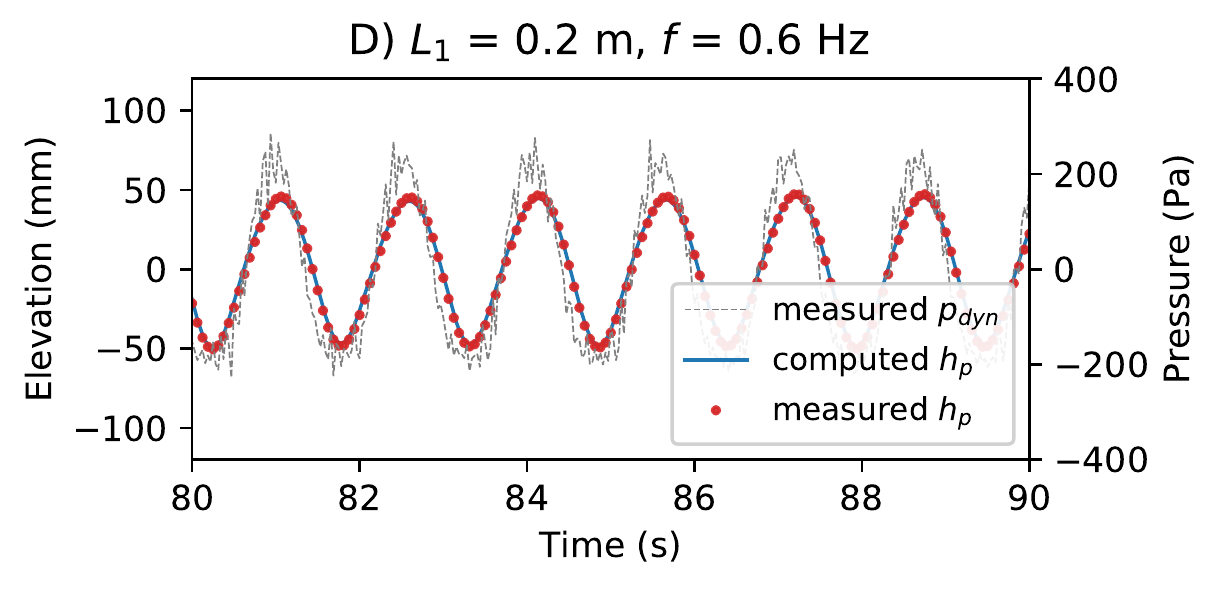}
\vfill
\includegraphics[width=6.7cm]{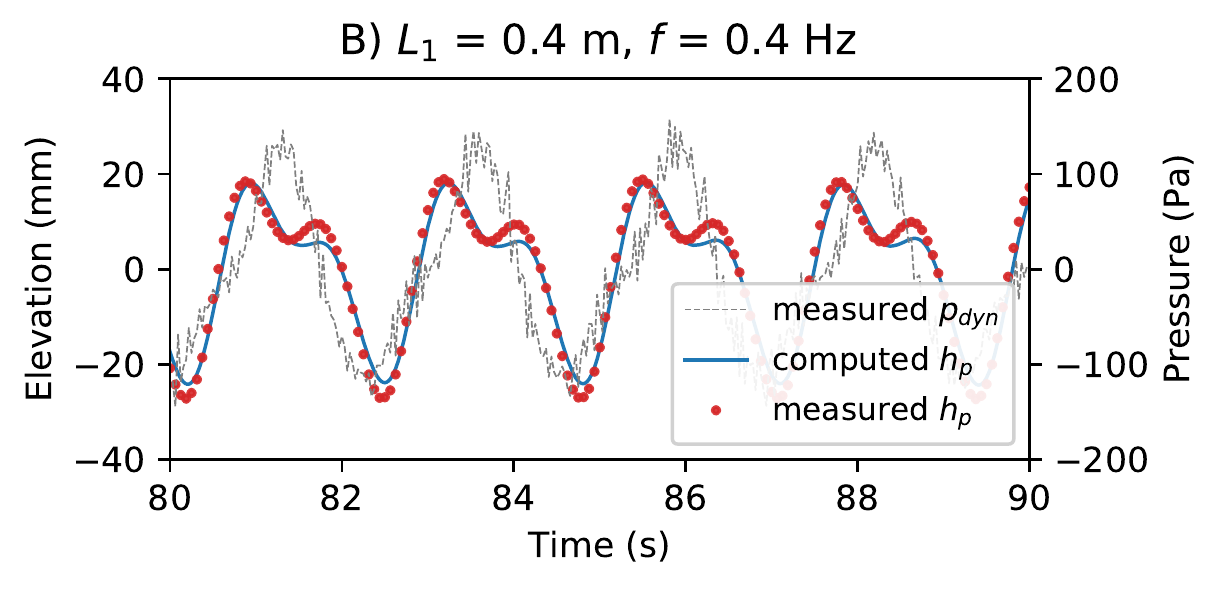}
\hfill
\includegraphics[width=6.7cm]{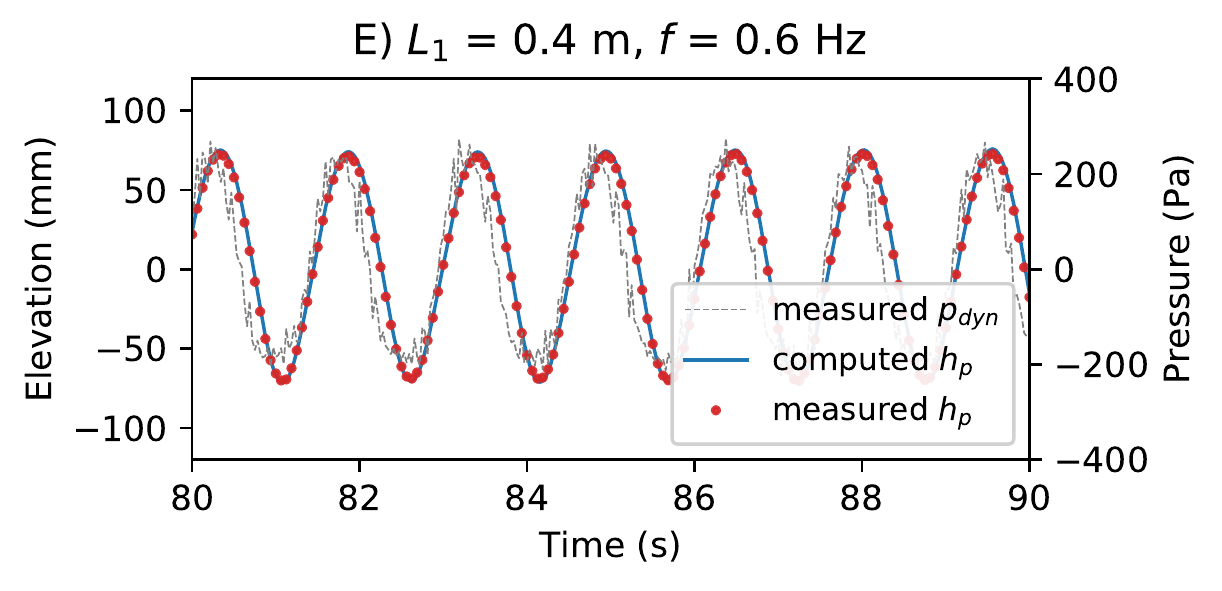}
\vfill
\includegraphics[width=6.7cm]{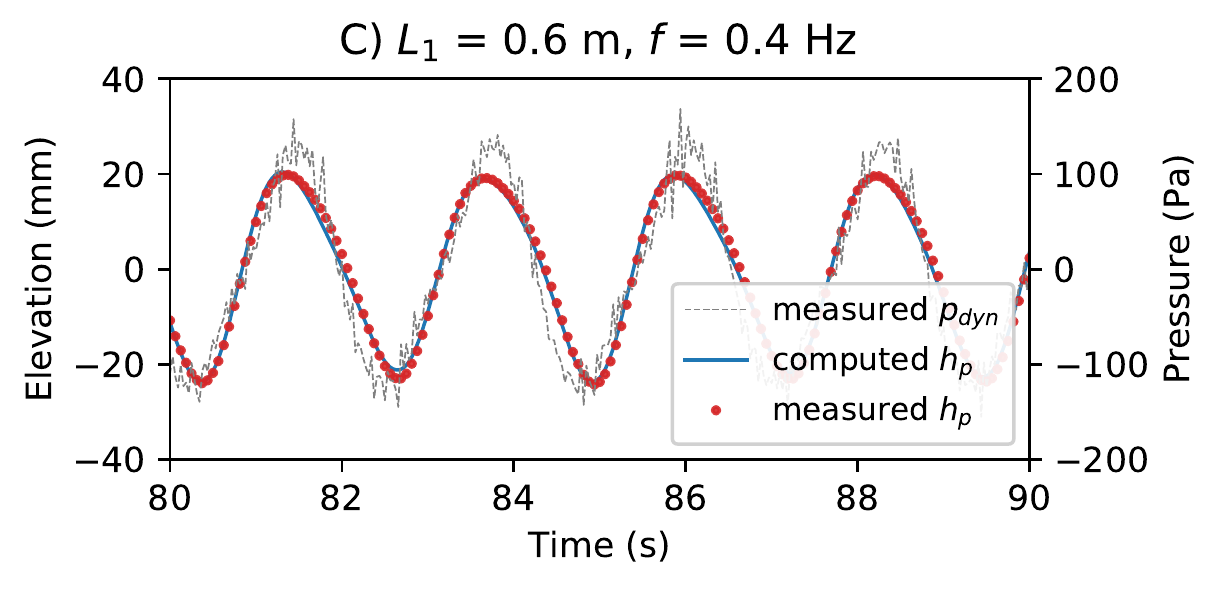}
\hfill
\includegraphics[width=6.7cm]{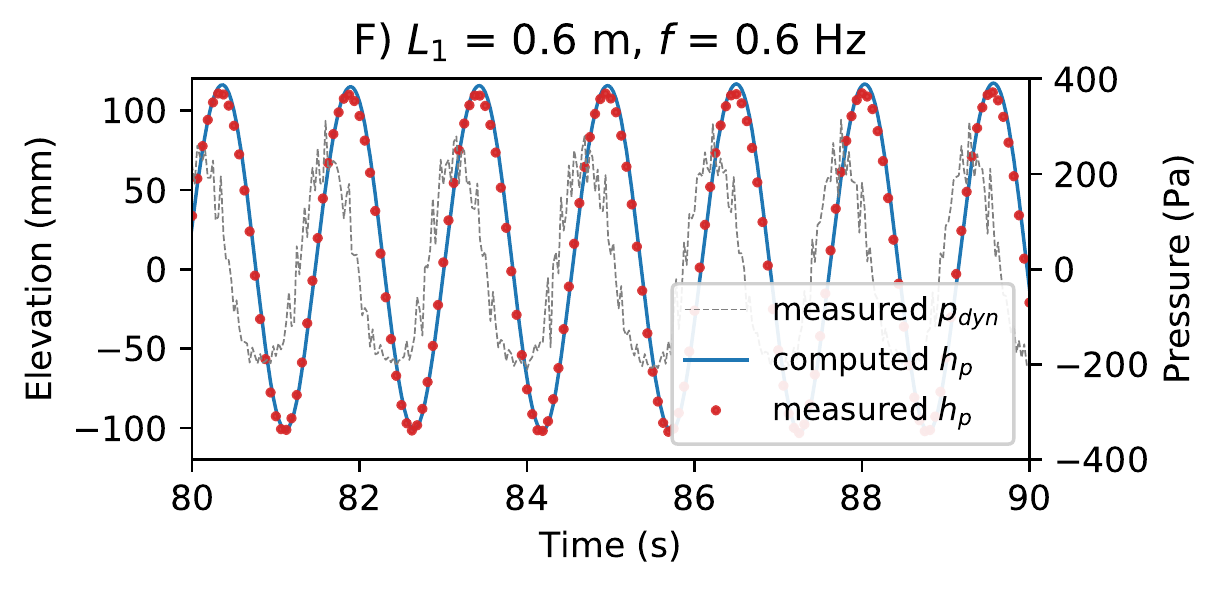}
\vfill
\caption{Comparison between measured and computed water level oscillations for the open-end pipe and for different pipe lengths and wave frequencies (10-sec excerpt)}
\label{fig:free_oscillations}
\end{figure}

Figure~\ref{fig:free_oscillations} presents a  10-sec time segment of water level oscillations inside the pipe system forced by two different wave conditions ($f = 0.4$ and 0.6 Hz) and for three different pipe geometries ($L_1=$0.2, 0.4 and 0.6 m). Although regular waves were generated by a wavemaker, because of the reflection from the vertical panel and wavemaker paddle, spurious waves appeared, which became noticeable at lower frequencies (Fig.~\ref{fig:free_oscillations}A, B). However, the water level oscillations were periodic. 
Both amplitude and phase computed by the proposed model are in excellent agreement with measured data. Note that the response of the water mass inside the pipes strongly depends on the geometry, namely the pipe length $L_1$.

Comparison of positive and negative amplitudes for all 15 considered scenarios are presented in Fig.~\ref{fig:comparison}A. Again, the agreement between the computed and measured water level amplitudes is satisfactory, with root mean square error RMSE = 5.1 mm.

\subsubsection{Water mass oscillations with compressed air}

To verify the complete numerical model (with special focus placed on the thermodynamics part of governing equations), the computed air pressure amplitudes were compared against measured values. The experiment set-up was the same as described in the previous subsection; however, in this case, the vertical pipe was closed by one of three different caps with small openings (Fig.~\ref{fig:exp_Scheme}B). Again, the wave pressure measured at the pipe inlet was imposed as a boundary condition for the numerical model. 

\begin{figure}[thbp]
	\center
	\includegraphics[width=6.7cm]{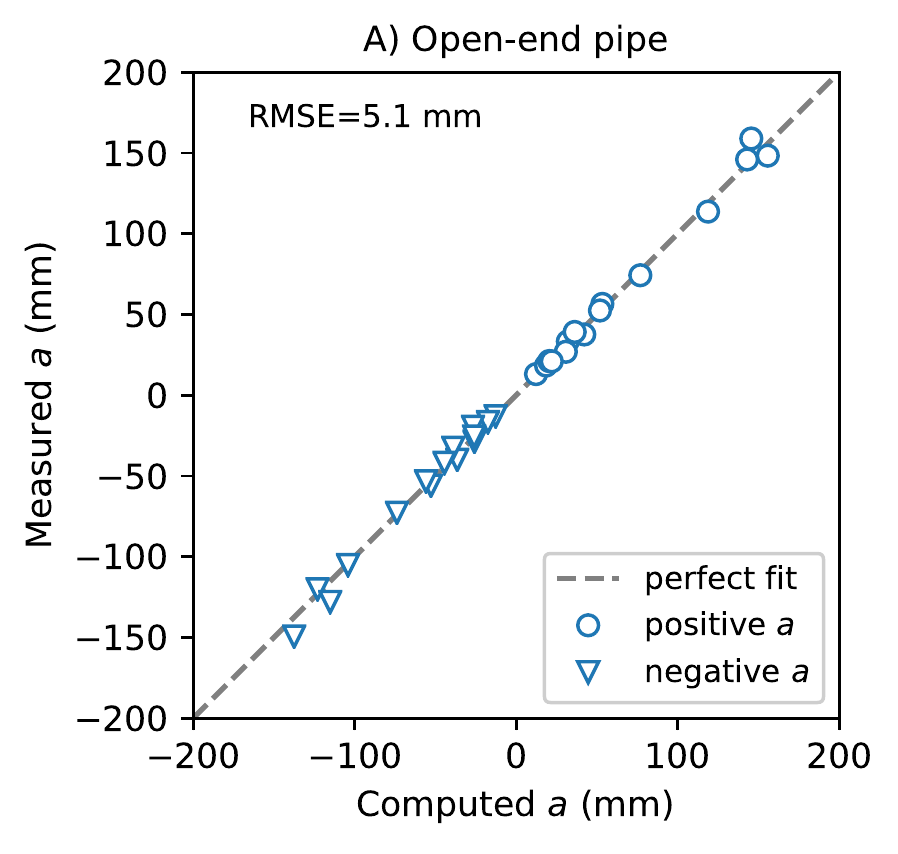}
	\hfill
	\includegraphics[width=6.7cm]{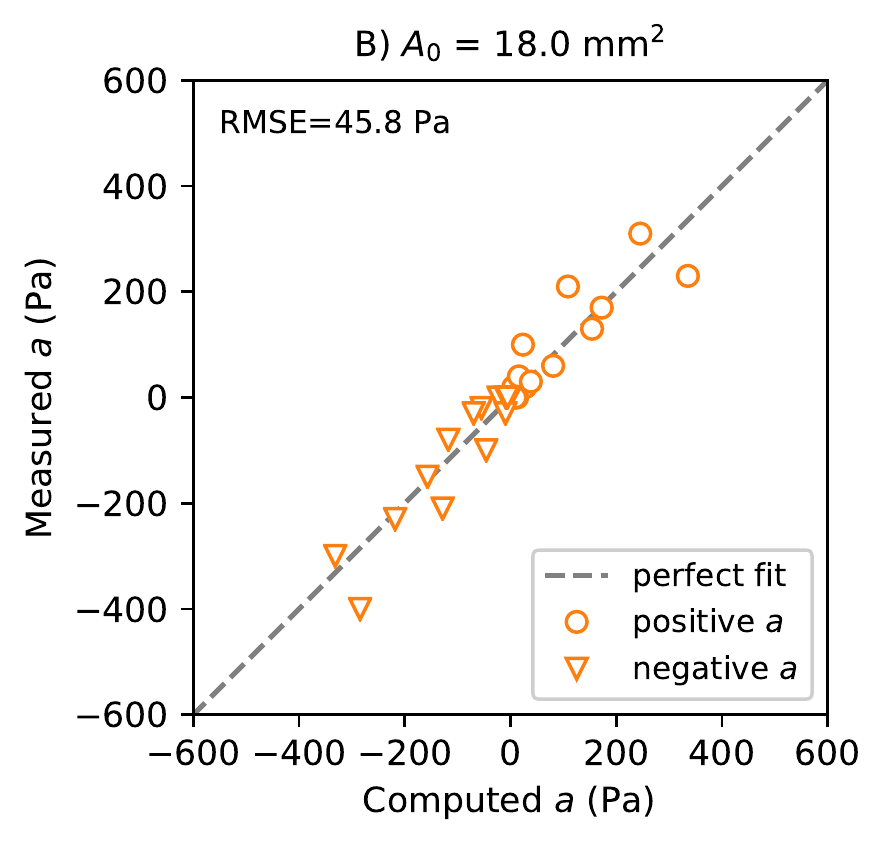}
	\vfill
	\includegraphics[width=6.7cm]{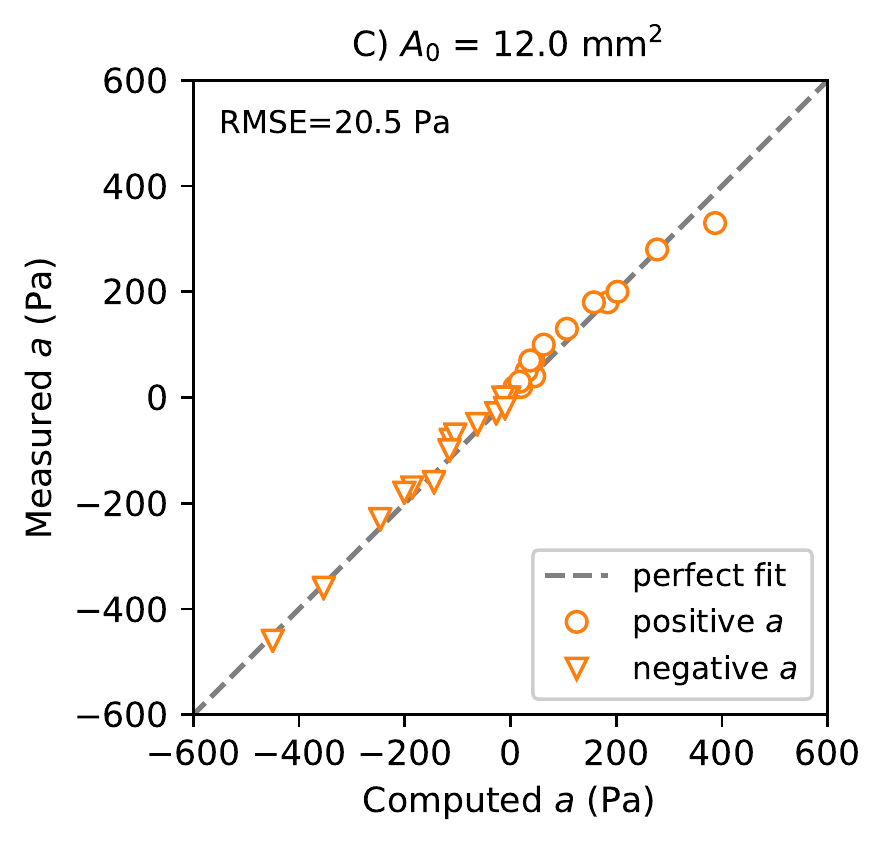}
	\hfill
	\includegraphics[width=6.7cm]{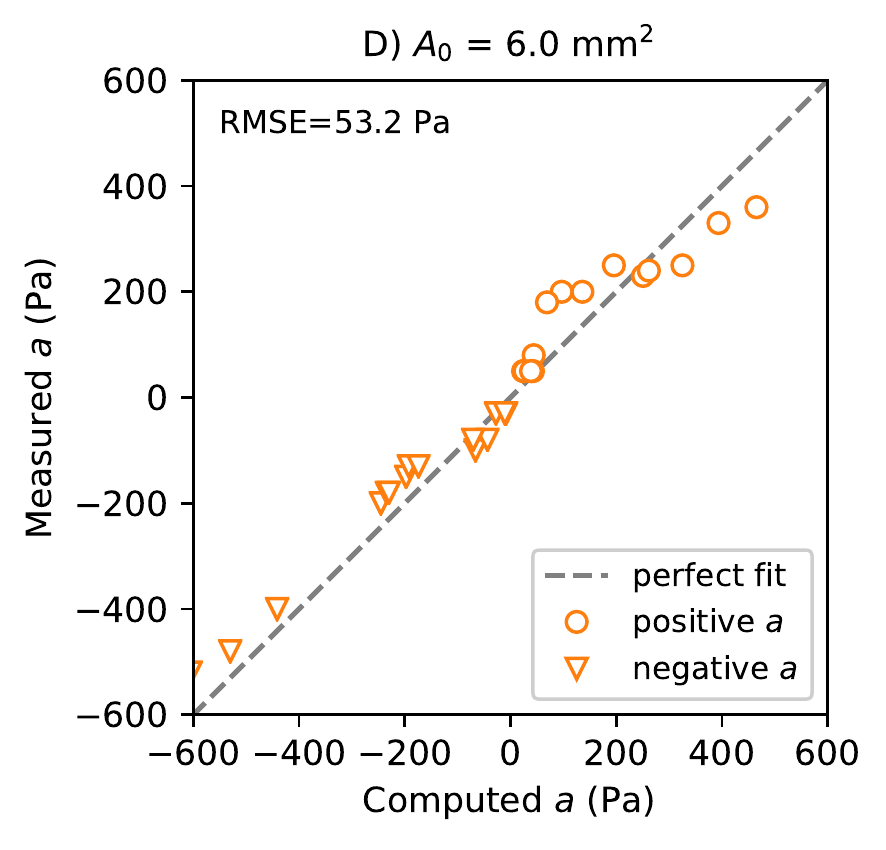}
	\vfill
	\caption{Comparison between measured and computed maximum amplitudes of: A) water level for open-end pipe, and B, C, D) air pressure for closed pipes with different orifice area $A_0$ = 18, 12 and 6 mm$^2$, respectively}
	\label{fig:comparison}
\end{figure}

Comparison of positive and negative pressure drop amplitudes for all 15 considered scenarios are presented in Fig.~\ref{fig:comparison}B, C and D for $A_0$ = 18, 12 and 6 mm$^2$, respectively. The best agreement was obtained for $A_0 = 12$ mm$^2$ (RMSE = 20.5 Pa), but the two remaining scenarios also show satisfactory agreement (RMSE = 45.8 and 53.2 Pa).

Note that the discharge coefficients were calibrated for each labium area by varying $C_d$ between 0.4 and 0.9 and finding the best fit with the experimental results. The values of $C_d$ =0.6, 0.64 and 0.7 were found for $A_0$ = 18, 12 and 6 mm$^2$, respectively. Similar values were obtained for OWC's orifice \citep{gervelas2011,iino2016} and air valves \citep{lingireddy2004,carlos2010}. It seems that either $C_d$ decreases with the orifice area, or that $C_d$ incorporates a correction factor for some unaccounted physical processes (such as turbulent effects), 
which become more prominent as the orifice area decreases.

\subsection{The Sea Organ analysis}

To demonstrate the model capabilities, internal oscillations in the Sea Organ forced by realistic wave conditions were simulated. The model set-up was defined similarly to the Sea Organ in Zadar. Unfortunately, exact dimensions are not publicly available, therefore the values were estimated from available design drawings (Fig.~\ref{fig:seaorgan_design}). One pipe from each segment was examined. Although pipes in each segment differ in size (according to the desired frequency of the sound, diameters $D_2$ and $D_3$ range from 50 to 125 mm) this difference has a negligible effect on the resulting internal oscillations in comparison to the overall dimensions of the pipe system.
Table~\ref{tab:seaorgan} shows middle pipe dimensions estimated from the design drawings for each of the seven segments.

\begin{table}[h]
	\small
	\centering
	\caption{Estimated dimensions of the middle pipe from each segment of the Sea Organ}
	\begin{tabular}{lrrrrrrr}
		\hline\noalign{\smallskip}
		\textbf{segment} & \textbf{1} & \textbf{2} & \textbf{3} & \textbf{4} & \textbf{5} & \textbf{6} & \textbf{7} \\
		\hline\noalign{\smallskip}
		$h_c$ (m asl) & 0.3  & 0.55  & 0.7  & 0.85  & 1.0  & 1.16  & 1.3 \\
		$h$ (m asl) & -0.2  & -0.2  & -0.2  & -0.2  & -0.2  & -0.2  & -0.2 \\
		$L_1$ (m) & 2.9   & 2.9   & 2.9   & 1     & 1     & 1     & 0.5 \\
		$L_2$ (m) & 5.26  & 4.26  & 3.60  & 4.26  & 3.60  & 2.55  & 2.20 \\
		$L_3$ (m) & 0.3   & 0.3   & 0.3   & 0.3   & 0.3   & 0.3   & 0.3 \\
		$D_1$ (m) & 0.3   & 0.3   & 0.3   & 0.3   & 0.3   & 0.3   & 0.3 \\
		$D_2$ (m) & 0.075   & 0.075   & 0.075   & 0.075   & 0.075   & 0.075   & 0.075 \\
		$D_3$ (m)  & 0.075   & 0.075   & 0.075   & 0.075   & 0.075   & 0.075   & 0.075 \\
		$\varphi$ ($^{\circ}$) & 20    & 25  & 30    & 25  & 30    & 45    & 55 \\
		$A_0$ (mm$^2$) & $112$ & $112$ & $112$ & $112$ & $112$ & $112$ & $112$ \\
		\hline\noalign{\smallskip}
	\end{tabular}%
	\label{tab:seaorgan}%
\end{table}%

\begin{figure}[thbp]
	\center
	\includegraphics[width=4.7cm]{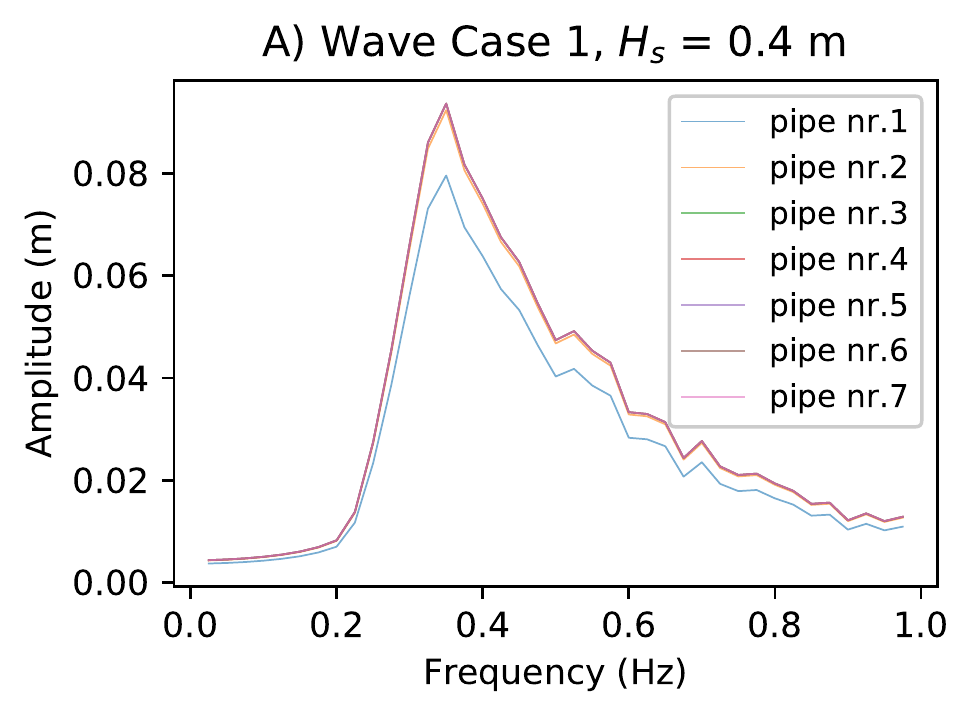}
	\hfill
	\includegraphics[width=8.6cm]{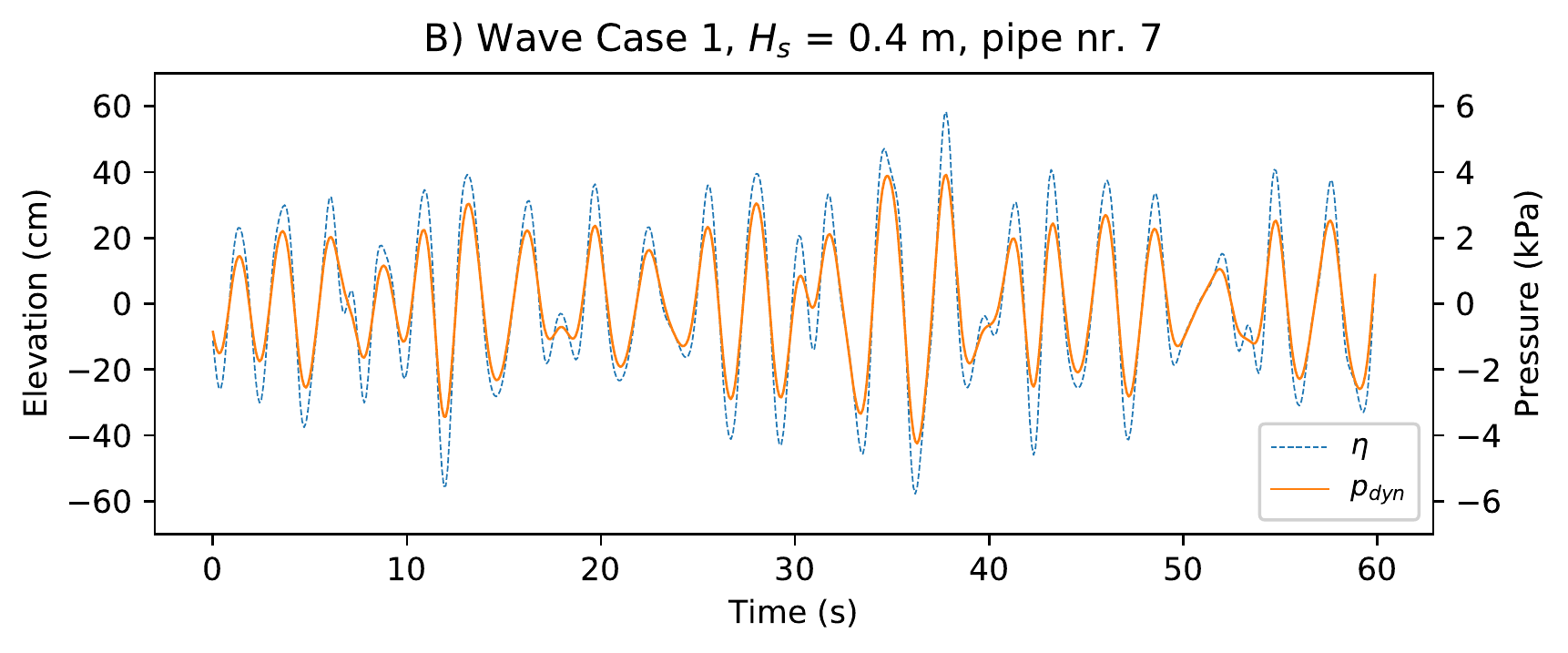}
	\vfill
	\includegraphics[width=4.7cm]{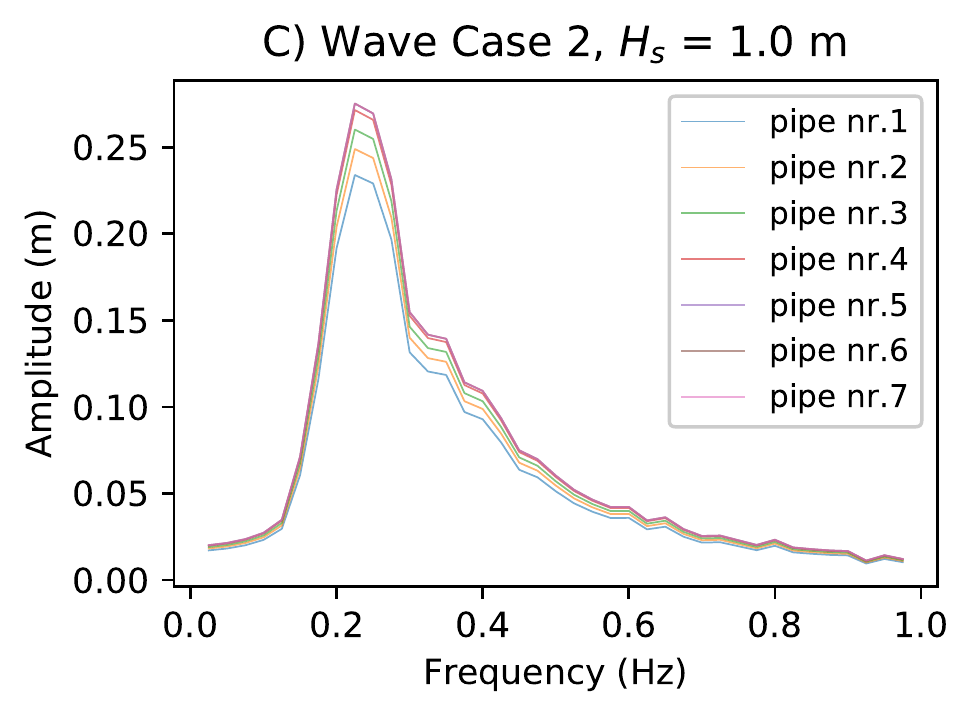}
	\hfill
	\includegraphics[width=8.6cm]{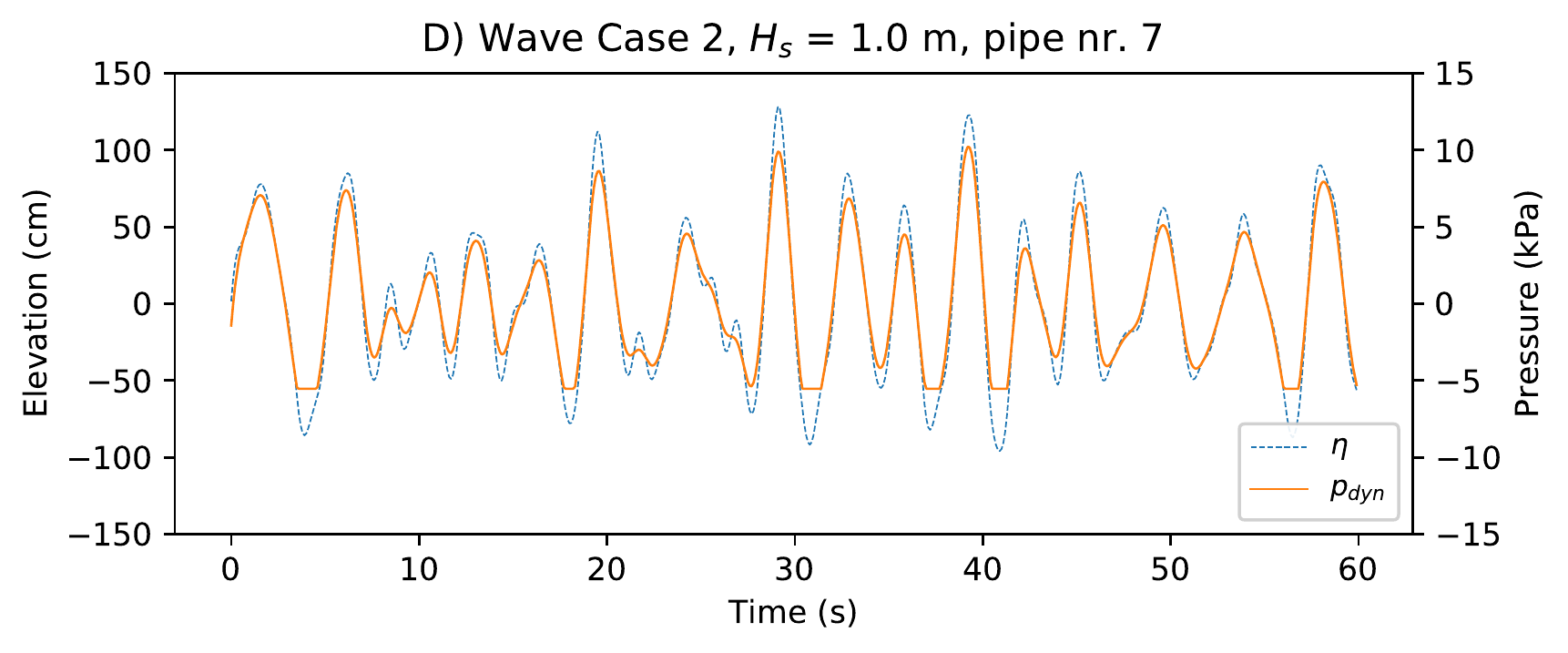}
	\vfill
	\caption{Generated wave amplitude spectrum and a 60-sec excerpt of water level with corresponding dynamic wave pressure at the pipe inlet for wave cases 1 and 2}
	\label{fig:waves2}
\end{figure}

The oscillations were forced by two irregular wave conditions generated from the T-spectrum \citep{tabain1997}: Case 1 was defined by $H_s = 0.4$ generated by a light northwest wind (Fig.~\ref{fig:waves2}A), whereas Case 2 was defined by $H_s = 1.0$ generated by a strong southeast wind (Fig.~\ref{fig:waves2}C). In both cases, still water level was set to +0.35 m asl. A 15-min wave field was simulated (Fig.~\ref{fig:waves2}B, D) and the corresponding water level and air pressure oscillations in the pipe system were computed. 

\begin{figure}[thbp]
	\center
	\includegraphics[width=6.7cm]{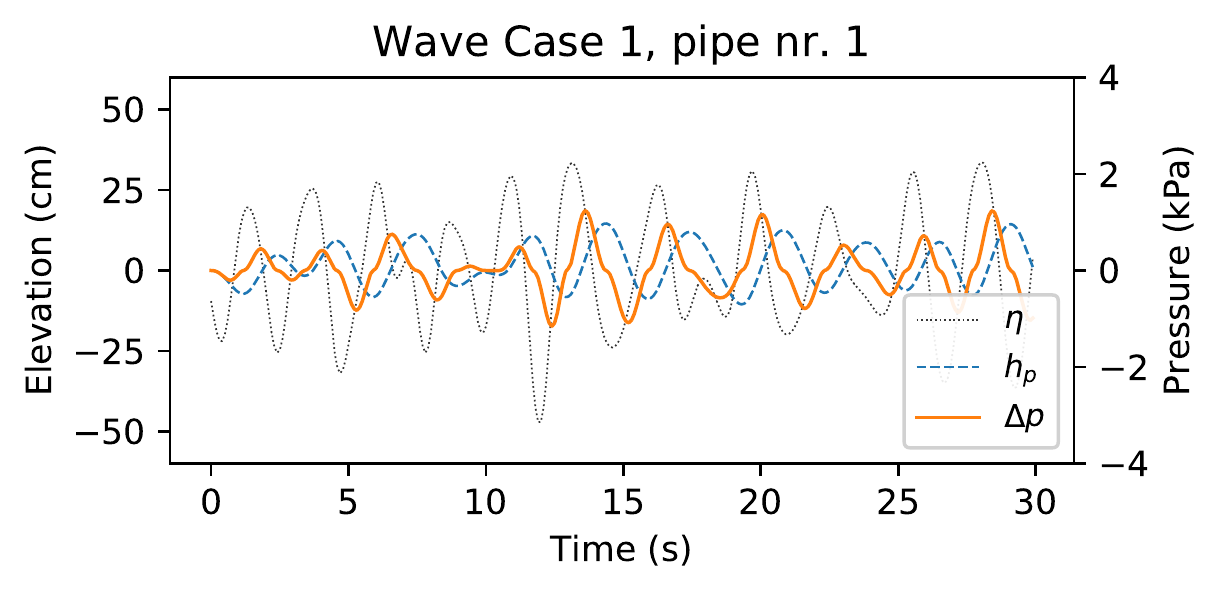}
	\hfill
	\includegraphics[width=6.7cm]{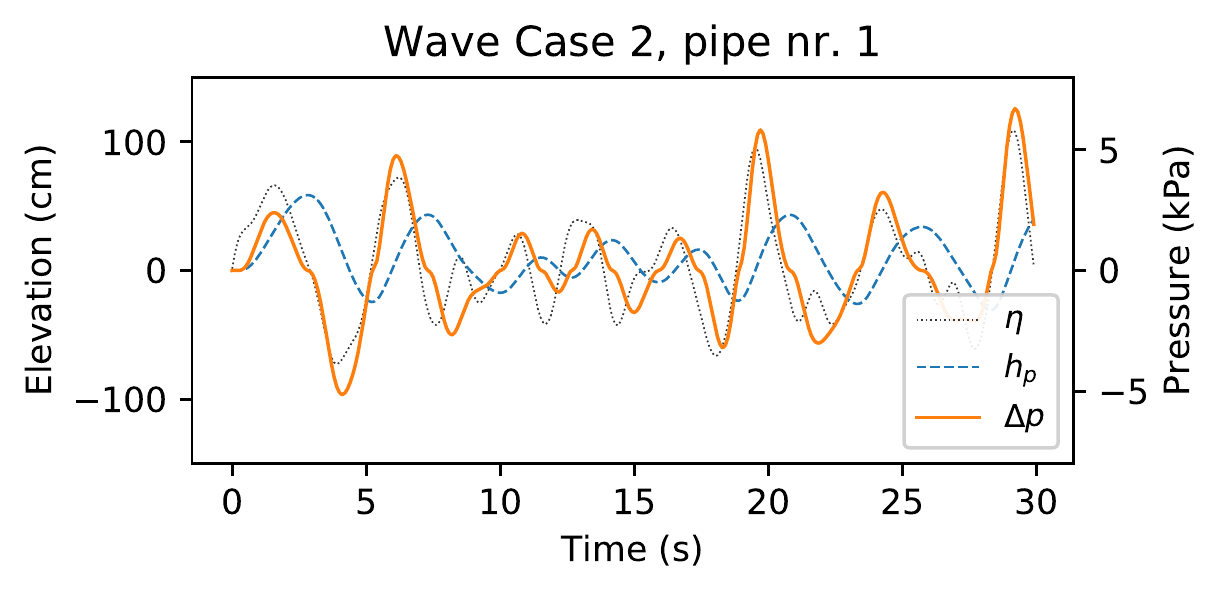}
	\vfill
	\includegraphics[width=6.7cm]{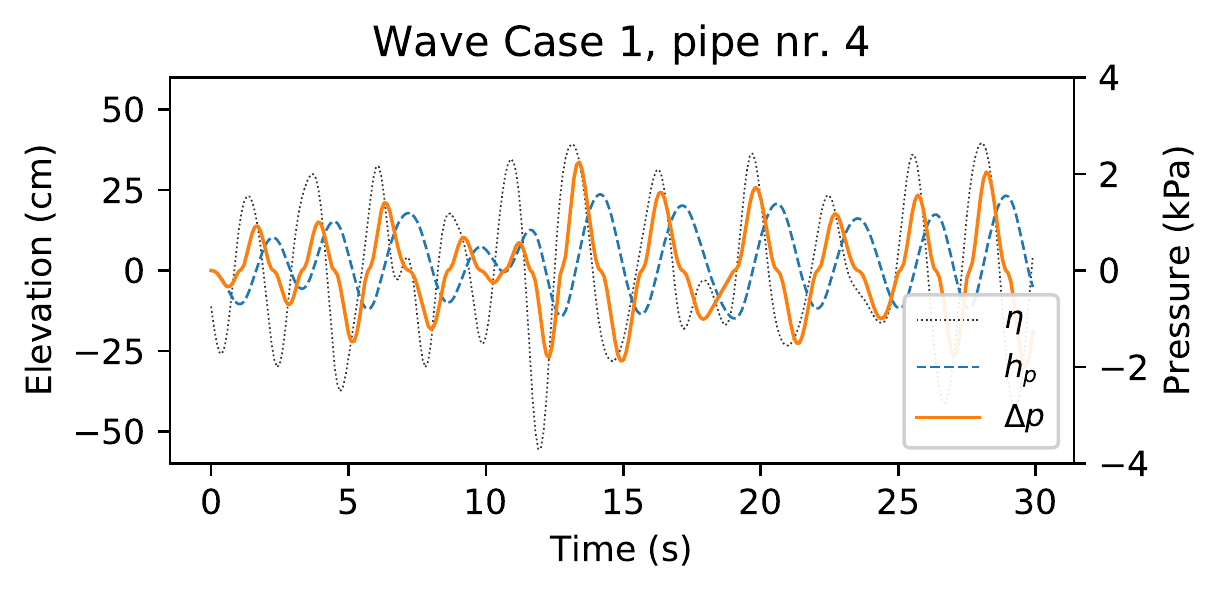}
	\hfill
	\includegraphics[width=6.7cm]{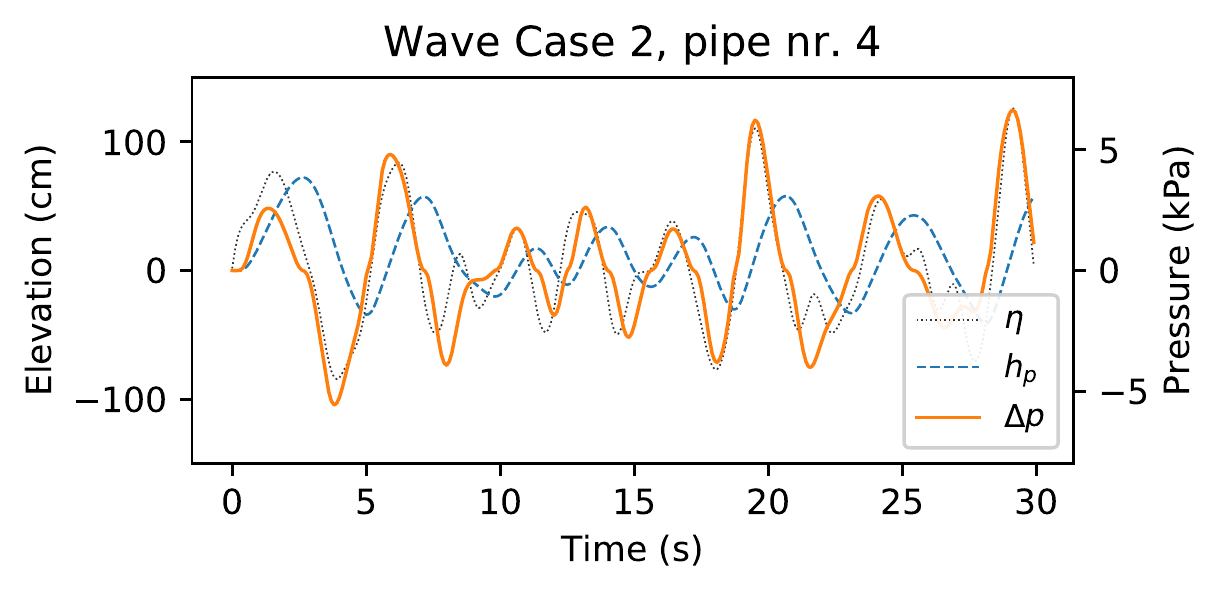}
	\vfill
	\includegraphics[width=6.7cm]{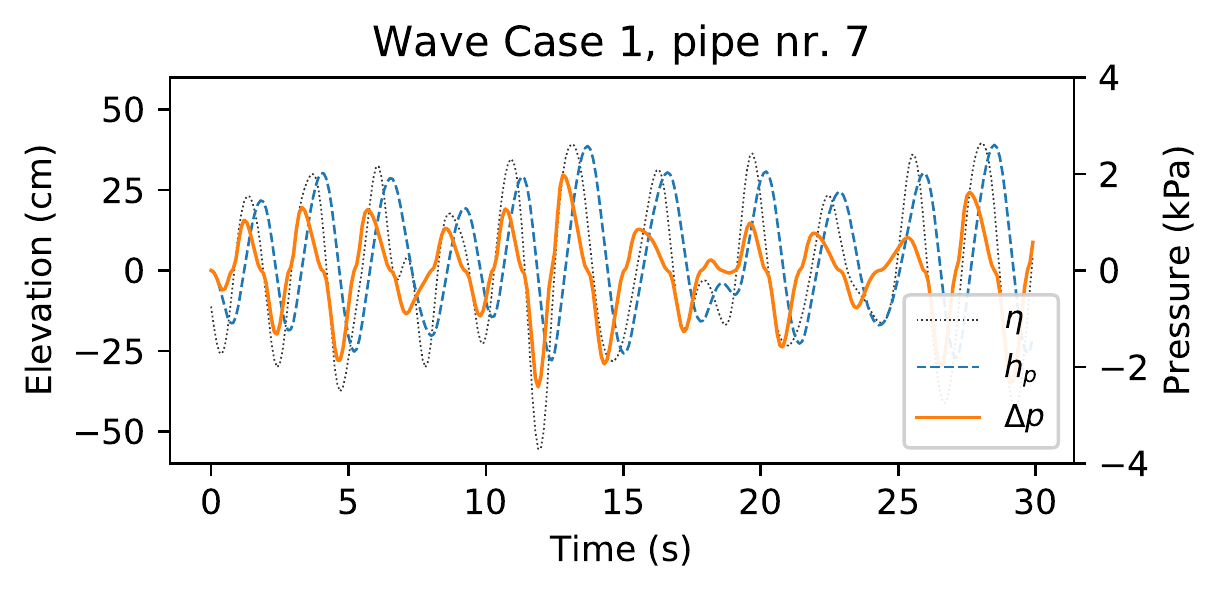}
	\hfill
	\includegraphics[width=6.7cm]{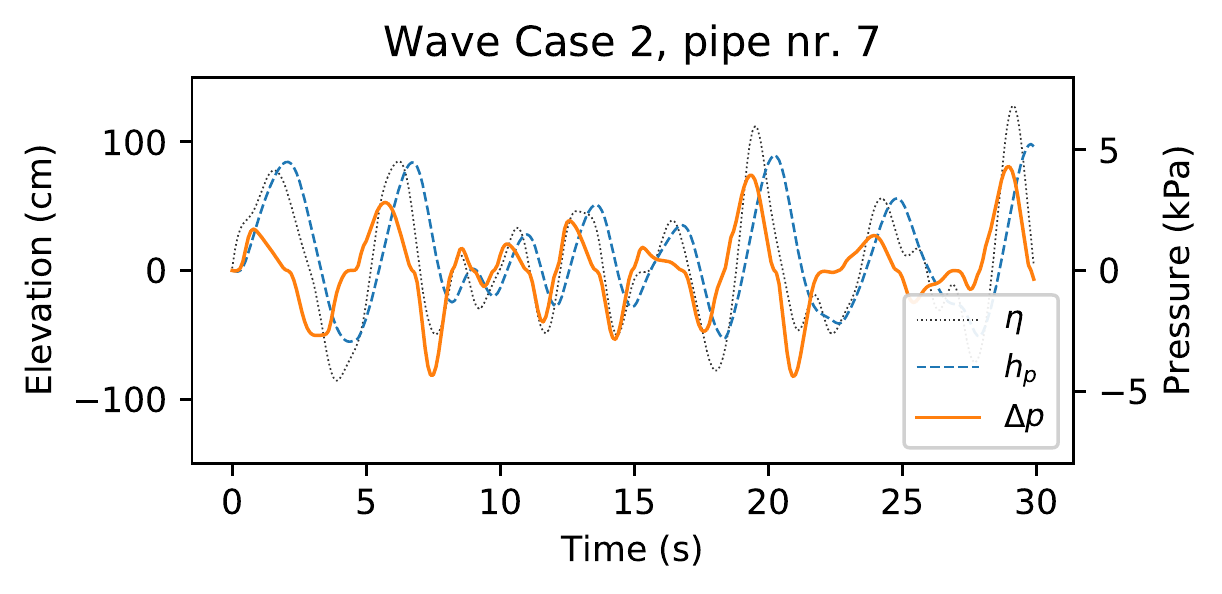}
	\caption{Comparison of sea surface elevations, internal water level elevations and air pressure drop oscillations in pipes 1, 4 and 7, for wave cases 1 and 2 (30-sec excerpt)}
	\label{fig:sea_organ_results}
\end{figure}

Figure~\ref{fig:sea_organ_results} shows the sea surface elevations, as well as the water level elevations and air pressure oscillations computed in pipes at three different segments (1, 4 and 7) for both wave scenarios. 
These results indicate that all segments are acoustically active, with the middle section providing the loudest sounds due to higher pressure. Clearly, higher waves generate a stronger response in the system and therefore internal oscillations are generally larger for $H_s = 1.0$ m than for $0.4$ m. It is important to emphasize that the response of internal oscillations differs not only in respect to waves but also between segments due to different pipe geometries.
For the first wave scenario, both the air pressure and water level responses in the first pipe are weaker in comparison to pipes 4 and 7. However, this is not the case for the second wave scenario, where the first pipe is equally responsive as the other two pipes. Furthermore, in both cases, air pressure oscillations are stronger in pipe 4 than in pipe 7. However, the opposite is true for water level elevations.

\begin{figure}[thbp]
	\center
	\includegraphics[width=6.7cm]{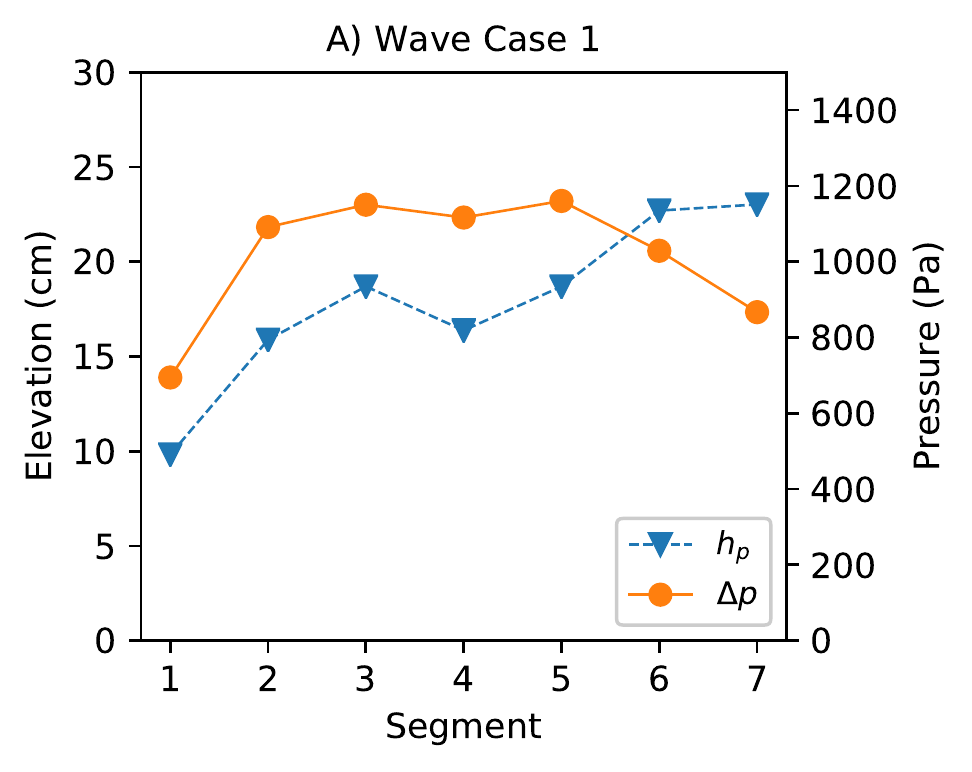}
	\hfill
	\includegraphics[width=6.7cm]{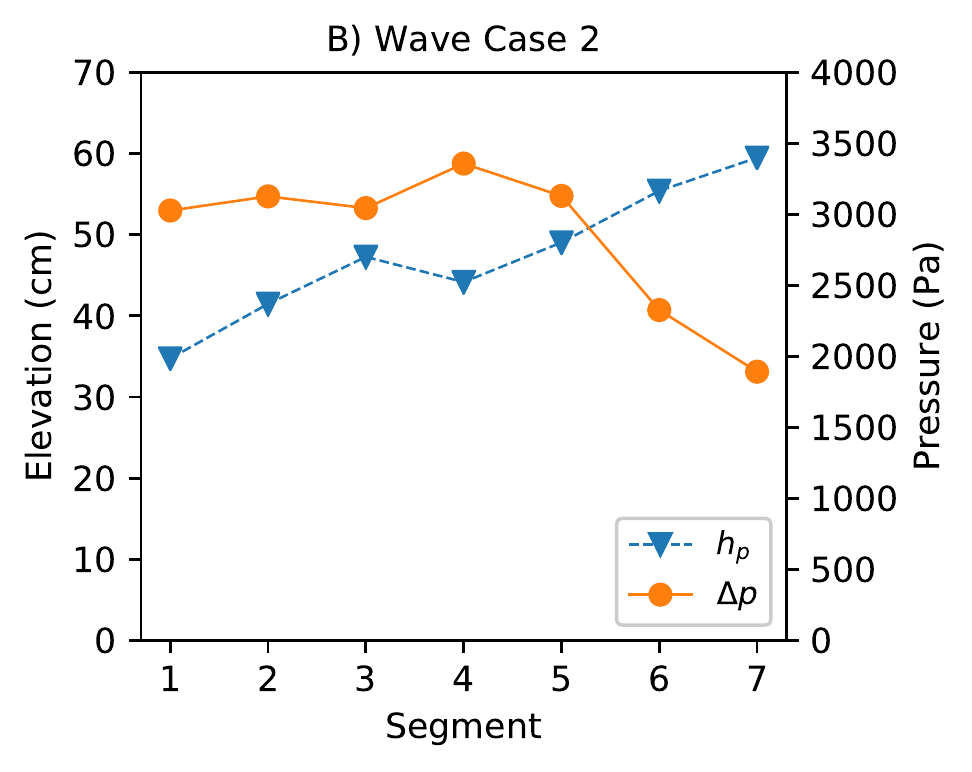}
	\caption{Mean amplitudes of water level elevation and air pressure drop in each segment of the Sea Organ for wave cases 1 and 2}
	\label{fig:sea_organ_sum_results}
\end{figure}

To illustrate the differences between pipes located in different segments, the mean amplitudes of internal oscillations are shown in Fig.~\ref{fig:sea_organ_sum_results}. A significant influence of the pipe geometry is noticeable; there is a clear discrepancy in both water level elevation and air pressure amplitudes between the segments. Furthermore, we can notice that the water mass in the same pipes responds quite differently to wave scenarios 1 and 2. Also, it seems that the pipe geometry has a different effect on the water level elevations than on the air pressure oscillations.

For the first wave scenario (Fig.~\ref{fig:sea_organ_sum_results}A), maximum air pressure amplitudes are found in pipes 3 and 5, and minimum in pipes 1 and 6. Water level amplitudes are lowest in the first and highest in the last two pipes. For the second wave scenario (Fig.~\ref{fig:sea_organ_sum_results}B), air pressure amplitudes are highest in the pipe 4 and smallest in the last two pipes. However, water level amplitudes show exactly the opposite. This result is in agreement with authors personal experiences from the Sea Organ in Zadar, where the sound from the first segment is quieter than the others or even non-existing during small waves, but for higher wave heights, sound from this segment can be heard quite loudly. 

\section{Discussion}

As mentioned before, internal processes in a sea organ could be considered similar to gravity-related processes in fixed OWC energy converters. It is commonly accepted that OWCs maximize the efficiency of wave energy extraction near resonance frequencies \citep{iino2016}. Let us now examine whether this is also true for the Sea Organ. Furthermore, we are interested in finding out how does the response of the sea organ system change with the pipe geometry and how to optimize the design of a sea organ pipe system.

\subsection{Natural frequency and resonance of the sea organ pipe system}

Internal oscillations in OWCs can be represented as mechanical single-degree-of-freedom systems (a rigid body) and their behaviour described by the equation of motion of the water column in forced and damped systems \citep{gervelas2011,iino2016}:
\begin{equation}
m\ddot{x} + c\dot{x} + kx = \varSigma F
\label{eq:equation_of_motion}
\end{equation}
where $x$ is the displacement of the water surface along the axis, $m$ is the mass of the water column, $c$ is the damping coefficient, $k$ is the restoring force and $\varSigma F$ is the sum of forces applied to the water mass. 

From Eq.~(\ref{eq:governing_eq}) it follows that Eq.~(\ref{eq:equation_of_motion}) is also applicable to sea organ internal oscillations, with $x = l_2$, where
\begin{align}
m(x) &= \rho \left( L_1 \frac{A_2}{A_1} + x\right)
\label{eq:motion_coeff1} \\
c(x) &= \frac{\beta \rho A_2 \lvert \dot{x} \rvert}{2} \\
k &= \rho g \sin \varphi \\
F(x) &= p_{wave} - \Delta p(x)
\label{eq:motion_coeff2}
\end{align}
Note, that from a strictly physical point of view, these coefficients represent the mass, damping coefficient, restoring gravity force and pressure forces per unit cross-section area. The pressure forces are a result of waves in front of the pipe inlet and compressed air in the acoustic pipe.

When the governing system is rewritten using Eqs.~(\ref{eq:motion_coeff1})-(\ref{eq:motion_coeff2}), the natural frequency of a water mass inside the sea-organ pipes can be expressed as \citep{harris2002}
\begin{equation}
f_n = 2 \pi \sqrt{\frac{k}{m}} = 2 \pi \sqrt{\frac{g \sin \varphi}{L_1 \frac{A_2}{A_1} + l_2}}.
\label{eq:natural_freq}
\end{equation}
From Eq. (\ref{eq:natural_freq}) we observe that the natural frequency changes with the inclination angle $\varphi$, length of the entry pipe $L_1$ (corrected by the corresponding cross-section area ratio) and length of the water column in the sloped pipe $l_2$. More, precisely, the natural frequency increases with $\varphi$ due to stronger gravity restoring force, but it decreases with $L_1$ and $l_2$ due to larger water mass. The latter relationship is expected and well known; however, the variability of the natural frequency with the inclination angle had been recognized and analysed only recently in OWCs \citep{iino2016}.

Furthermore, since the governing system includes viscous damping, the natural frequency should be corrected as follows \citep{harris2002}:
\begin{equation}
f_{d} = f_{n} \left(1 - \zeta^2\right)^{1/2},
\label{eq:damped_freq}
\end{equation}
where $\zeta=c/c_c$ is the damping ratio, and $c_c = 2\sqrt{km}$ is the critical damping. Finally, maximum displacement response is expected near the displacement resonance frequency, which is defied as \citep{harris2002}
\begin{equation}
f_{r} = f_{n} \left(1 - 2\zeta^2\right)^{1/2}.
\label{eq:resonance_freq}
\end{equation}

Table \ref{tab:natural_freq} shows all three frequencies for each segment of the Sea Organ. Natural frequency is computed by Eq.~(\ref{eq:natural_freq}), $f_d$ is obtained by a numerical analysis of the water level oscillations in the sea-organ pipes (with $p_{wave}(t) = const.$ and an initial increase of the water level in the sloped pipe), $\zeta$ is computed by Eq.~(\ref{eq:damped_freq}), and $f_r$ is then estimated from Eq.~(\ref{eq:resonance_freq}).

\begin{table}[htbp]
	\small
	\centering
	\caption{Natural, damped and resonance frequencies for each segment of the Sea Organ}
	\begin{tabular}{rlllllll}
		\hline\noalign{\smallskip}
		segment & 1     & 2     & 3     & 4     & 5     & 6     & 7 \\ 
		\hline\noalign{\smallskip}
		$f_n$ (Hz)   & 0.218 & 0.266 & 0.311 & 0.277 & 0.327 & 0.457 & 0.538 \\
		$f_d$ (Hz)   & 0.216 & 0.265 & 0.310 & 0.276 & 0.325 & 0.453 & 0.533 \\
		$\zeta$ (-)  & 0.115 & 0.101 & 0.103 & 0.087 & 0.116 & 0.137 & 0.140 \\
		$f_r$ (Hz)   & 0.215 & 0.263 & 0.308 & 0.275 & 0.323 & 0.449 & 0.528 \\
		\hline\noalign{\smallskip}
	\end{tabular}%
	\label{tab:natural_freq}%
\end{table}%

Table \ref{tab:natural_freq} shows that the viscous damping is well under the critical damping coefficient $c_c$. Therefore, for each pipe, all three natural frequencies are very similar. However, natural frequencies differ between the segments; the first pipe has the lowest natural frequency $f_n = 0.218$ Hz, whereas the last pipe has the highest frequency $f_n = 0.538$ Hz. 

If we consider the first wave scenario, characterized by the peak frequency $f_p = 0.34$ Hz (Fig.~\ref{fig:waves2}A), sea organ efficiency should be more prominent in pipes 3 and 5 due to similar values of $f_r$. Fig.~\ref{fig:sea_organ_sum_results}A suggests that pressures in pipes 3 and 5 indeed have the highest mean amplitude; however, water level elevations in the same pipes are lower than in pipes 6 and 7. Moreover, for the second wave scenario, characterized by the peak frequency $f_p = 0.22$ Hz (Fig.~\ref{fig:waves2}C), sea organ efficiency should be more prominent in the first pipe due to similar $f_r$. However, Fig.~\ref{fig:sea_organ_sum_results}B suggests that the highest pressure amplitudes occur in pipe 4, and highest water level amplitudes in pipe 7.

These results indicate that the maximal values of water level elevations and air pressures in a sea-organ are a result of several different effects and that they cannot be predicted only by the resonance. First of all, although the same waves are generated in front of the Sea Organ wall, not all segments are forced by the same wave pressure. In addition to inlet depth (which is the same for all segments in this example), wave pressure is directly linked to the local sea surface elevations which are influenced by the reflected waves. Since the crest height differs between the segments (Table \ref{tab:seaorgan}), so does the reflection coefficient and the resulting wave pressures at each pipe inlet. In other words, lower oscillation amplitudes in the first pipes are partially the result of lower sea surface elevations. Furthermore, Fig.~\ref{fig:sea_organ_sum_results} shows the resulting amplitudes of water level elevations $h_p$; however, the second pipe is inclined, hence water level displacement in the pipe axis direction $l_2$ should give a more realistic information on the effect of resonance. 

\begin{figure}[thbp]
	\center
	\includegraphics[width=6.7cm]{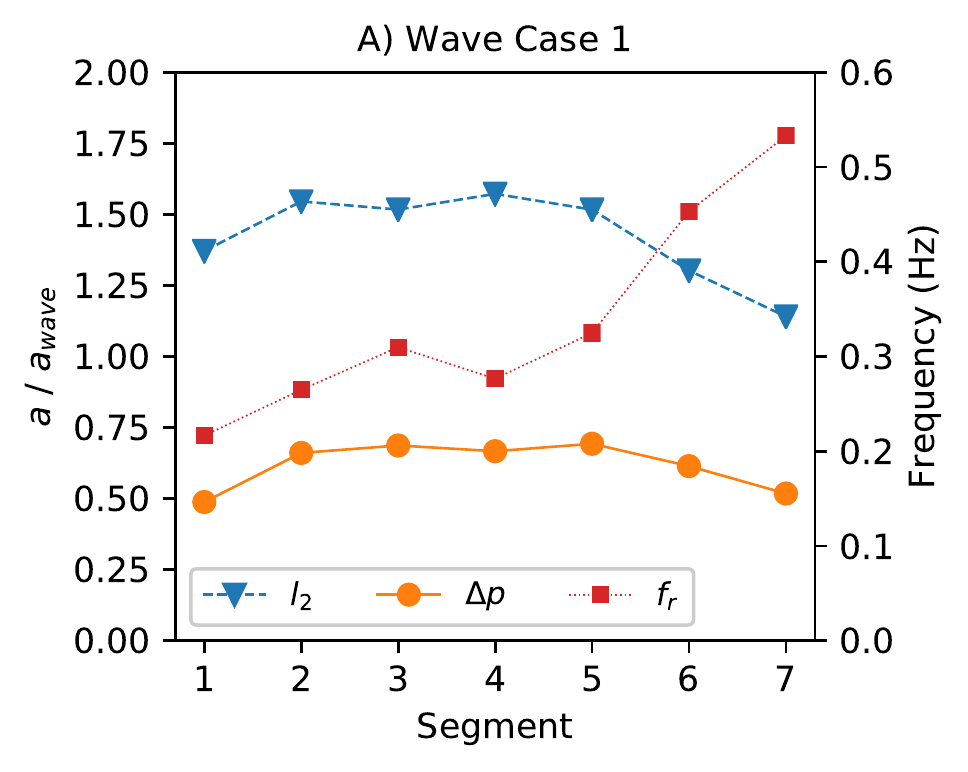}
	\hfill
	\includegraphics[width=6.7cm]{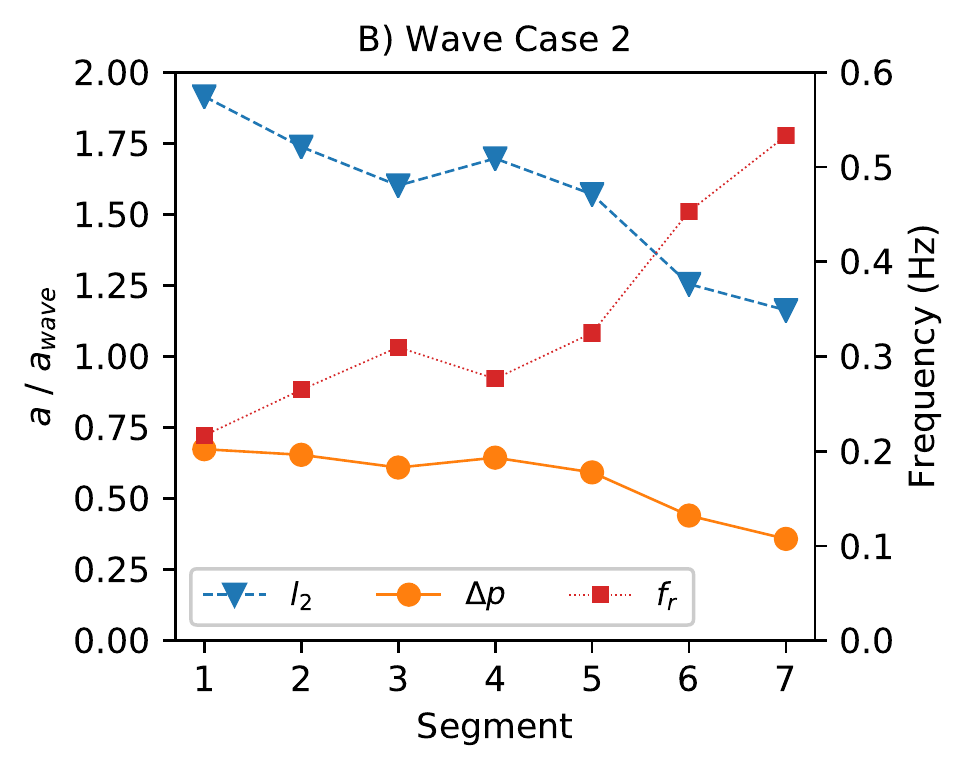}
	\caption{Relative mean amplitudes of water level and air pressure drop in each segment of the Sea Organ for wave cases 1 and 2}
	\label{fig:ratios_results}
\end{figure}

To compensate for these additional effects and focus only on the resonance, the same results are presented again in Fig.~\ref{fig:ratios_results}, which shows the mean amplitude ratio $a/a_{wave}$ of the water level displacement $l_2$ to sea surface elevation $\eta$ and the air pressure drop $\Delta p$ to dynamic wave pressure $p_{dyn}$. The corresponding resonance frequencies for each segment (Table~\ref{tab:natural_freq}) are also illustrated for clarity.

Fig.~\ref{fig:ratios_results} confirms that the strongest response of internal oscillations is in fact the result of resonance. For the first wave scenario, the highest water level displacement and air pressure amplitude ratio is found in pipes 2-5 that have $f_r = 0.263 - 0.323$ Hz, which are closest to $f_p = 0.34$ Hz. Similarly, for the second wave scenario, the highest $a/a_{wave}$ are computed at the first pipe that has $f_r = 0.215$ Hz, which is closest to $h_p = 0.22$ Hz.

\section{Conclusion and recommendation}

The aim of this study was to develop a numerical model for predicting the water level and air pressure oscillations in a sea organ forced by regular and irregular waves. The model was derived by coupling hydrodynamic and thermodynamic equations under the assumption of incompressible water flow, isentropic gas processes, and negligible turbulent contributions.

Although the model is relatively simple, it has shown satisfactory agreement with small-scale measurements. The main advantage of the proposed model is its computational speed, especially when compared to more advanced numerical models, such as the family of multiphase 3D CFD models. Since the model gives a good insight in the internal physical processes, it should be a valuable tool, not only in the preliminary design of similar acoustical structures but also in the design of fixed OWC energy converters with complex internal geometry.

From the numerical analysis of the Sea Organ in Zadar, we found that internal oscillations respond quite differently depending on the wave conditions. As expected, both water level and air pressures increase with the wave height. Differences in internal oscillations between segments due to different geometries are also noticeable. The resulting water level and air pressure oscillations are most sensitive with respect to the inclination angle and length of the pipes. Furthermore, we confirmed that the sea-organ is most efficient when the resonance frequencies of the water mass are close to peak wave frequencies. However, inclination angle must also be considered when water elevation is considered; water level displacements in the inclined pipe axis direction does not necessarily coincide with maximum water level elevations. This also has some significance when inclined OWC energy converters are considered.

In comparison to OWCs, where the only concern is maximizing the efficiency of energy extraction, in a sea organ, there are several objectives. First of all, we are primarily interested in the air pressure drop which directly governs the sound amplification, but the water level elevations are also relevant with respect to the structural safety and reliability. Additionally, these goals can be quite diverse depending on the wave conditions. To be more precise, during small waves, the main aim is to maximize the efficiency of all sea-organ pipes; however, for large waves, the goal is to minimize its efficiency in order to prevent the extreme sound loudness and water intrusion into the acoustical pipes which can damage finely tuned elements.

Finally, the recommendations for the design of sea organ pipes from a hydraulic perspective can be summarized as follows:
\begin{itemize}
	\item The geometry of the sea organ pipes should be designed so that the resonance frequency of the internal water mass is close to the peak frequency for smaller waves and far from the peak frequency for larger waves to ensure optimal sound loudness under all wave conditions. Precise definition of \textit{small} and \textit{large} waves depends on the local wave climate and the elevations of  acoustical pipes.
	\item To accomplish the first goal, the resonance frequency can be increased by using shorter pipe lengths and steeper pipe inclination angles, and \textit{vice versa}.
	\item Resonance analysis gives a good estimate of maximum air pressures in the acoustical pipe; however, a numerical time-series analysis must also be performed in order to examine the water level displacements and prevent a possible water intrusion into the acoustical pipe.
\end{itemize}

\appendix
\section*{Acknowledgements}
This work has been supported in part by Ministry of Science, Education and Sports of the Republic of Croatia under the project \textit{Research Infrastructure for Campus-based Laboratories at the University of Rijeka}, number RC.2.2.06-0001, which was co-funded from the European Regional Development Fund (ERDF).

\section*{References}
\small
\bibliographystyle{elsarticle-harv} 
\bibliography{references}

\end{document}